\documentclass[11pt,letterpaper]{article}
\usepackage{jheppub}
\usepackage{times}  

\title{Conserved cosmological perturbation in Galileon models}

\author[a,b,c]{Xian Gao}

\affiliation[a]{\href{http://www.apc.univ-paris7.fr/APC_CS/en}{Astroparticule et Cosmologie (APC)}, UMR 7164-CNRS, Universit\'{e} Denis Diderot-Paris 7,\\
        10 rue Alice Domon et L\'{e}onie Duquet, 75205 Paris, France}
\affiliation[b]{\href{http://www.lpt.ens.fr/?lang=en}{Laboratoire de Physique Th\'{e}orique, \'{E}cole Normale Sup\'{e}rieure (LPTENS)},\\
    24 rue Lhomond, 75231 Paris, France}
\affiliation[c]{\href{http://www.iap.fr/english/}{Institut d'Astrophysique de Paris (IAP)}, UMR 7095-CNRS, Universit\'{e} Pierre et Marie Curie-Paris 6,\\
98bis Boulevard Arago, 75014 Paris, France.}

\emailAdd{xgao@apc.univ-paris7.fr}

\date{\today}


\abstract{
    We prove the existence of a fully nonlinear conserved curvature perturbation on large scales in Galileon-type scalar field models in two approaches. The first approach is based on the conservation of energy-momentum tensor of the Galileon field, which is also the familiar approach in understanding the conservation in $k$-essence or perfect fluid models. We show that the fluid corresponding to the Galileon field becomes perfect and barotropic on large scales, which is responsible to the conservation. The difference from $k$-essence model is that, besides the energy-momentum conservation, the Einstein equation must be employed in order to complete the proof of barotropy. In the second approach, we derive the fully non-perturbative action for the curvature perturbation $\zeta$ in Galileon models on large scales, and argue that $\zeta=const$ is indeed an exact solution   on large scales. This conservation of curvature perturbation is important since it relates the later and  the primordial universe.
}

\begin{document}

\maketitle

\section{Introduction and motivation}

Inflation \cite{inflation} is one of the most exciting and successful ideas in modern cosmology. Over the years, inflation has become so popular because of its prediction of a nearly scale-invariant primordial
density perturbation, which was generated and frozen during inflation
to seed wrinkles in the Cosmic Microwave Background(CMB) \cite{Larson:2010gs} and today's Large-scale Structure
(LSS).

The most popular and widely studied inflationary models are mostly based on scalar fields \cite{inf_model}. Among them, $k$-essencial scalar models \cite{kess} have attracted much attention due to their broad generality, since which can be viewed as the most general scalar field theories with Lagrangians containing derivatives up to the \emph{first} order: $\mathcal{L}=\mathcal{L}(\phi,\nabla\phi)$. Indeed, this is the ``normal" construction of field theories. Nevertheless, one can make a step further to consider higher order derivative theories. However, these theories are often suffered with unexpected degree(s) of freedom such as ghost(s), breakdown of causality as
well as instability problems. This fact makes the construction of scalar field models beyond the first-order derivative be not trivial.

In the study of  DGP theory \cite{Dvali:2000hr} as well as some consistent
theories of massive gravity \cite{dgpmass}, a term $(\nabla\phi)^2\square\phi$ arises in the decoupling limit, which yields a second-order equation of motion and thus prevents possible extra degree of freedom and ghost. This is the simplest example of a higher-order derivative scalar field Lagrangian which still keeps the equation of motion as second-order{\footnote{Actually, General Relativity is another elegant example, since the Ricci scalar $R$ contains second order derivatives of the metric, but the resulting Einstein tensor in the equation of motion is also up to the second order derivatives. This is because the second derivatives of the metric enter $R$ \emph{linearly}.}} (we refer to this property as ``second-order preserving").
Inspired by this result, the so-called ``Galileon" was proposed (and also named) in \cite{Nicolis:2008in} in flat background and extended in \cite{cdcov} in general background, which introduced more general ``second-order preserving" terms  which are \emph{nonlinear} in second-order derivatives.

Historically, the most general scalar-tensor theories with second-order equations of motion in four dimensions was firstly explored by Hordeski in 1974 \cite{Horndeski}. This idea --- keeping the equation of motion as second order --- was further developed independently and extended in \cite{Deffayet:2011gz} following the concept of ``Galileon", in which the most general scalar field theories which we refer to as the ``generalized Galileons", whose Lagrangian contains derivatives up to the second order $\mathcal{L}=\mathcal{L}(\phi,\nabla\phi,\nabla\nabla\phi)$, but still keeps the equations of motion as second order as well as \emph{linear} in the second-order temporal derivative $\ddot{\phi}$ in arbitrary dimensions, was constructed. It can be shown that Horndeski's theory is included in and equivalent to the ``generalized Galileons" \cite{Deffayet:2011gz,kyy_gg}.
Special cases of the ``generalized Galileons" can also be realized as probing brane embedded in a higher dimensional
spacetime \cite{brane_ga}.
The ``generalized Galileons", opens up new windows in the study of cosmology and inflationary model construction, which have not been explored before. Since the invention of Galileons, a number
of applications have been made to various contexts in cosmology \cite{app_galileon,kyy_prl,kyy_gg}.

Before applying Galileon models to various aspects of cosmology, however, there is an important issue should be investigated: the existence (or non-existence) of a conserved curvature perturbation on super-Hubble scales.
As in the mostly accepted picture of cosmology, the evolution of cosmological perturbations can be divided into three stages: 1) generation during the early period of inflation as quantum fluctuations, 2) exiting the Hubble radius during inflation and becoming super-Hubble classical perturbations, 3) finally re-entering the Hubble radius in the later radiation/matter dominated era.
CMB and LSS formed in the third stage, in which we need to set the ``initial conditions" for their evolution. These initial conditions are typically given by simply the primordial information, which is evaluated in the first stage. What ensures this procedure is that, the curvature perturbation is conserved on super-Hubble scales, which allows us to relate the primordial and the later perturbations directly.

The previous understanding of conservation of curvature perturbation highly relies on the study of  perfect fluid or $k$-essencial scalar field. It is well-known that for perfect fluid, there exists a fully nonlinear curvature perturbation, which is conserved on large scales to fully nonlinear perturbative orders, if the pressure is a function of the energy density, i.e. if the fluid is \emph{barotropic} \cite{cons_gen,geocons}. Equivalently, the corresponding perturbation is called \emph{adiabatic}. On the other hand, for perfect fluid or $k$-essencial scalar field, scalar perturbation always becomes adiabatic on large scales \cite{Gordon:2000hv,Christopherson:2008ry,Langlois:2008mn}. This fact explains the existence of a conserved curvature perturbation in such models.

From the point of view of $k$-essence \cite{kess}, the Galileon field is quite strange: it is a model of a single scalar field, whereas its energy-momentum tensor takes the form of an \emph{imperfect} fluid \cite{Deffayet:2010qz,Pujolas:2011he}. This is contrary to the case of $k$-essence scalar, whose energy-momentum tensor can always be cast into a perfect fluid form in its own comoving frame.
Furthermore, as we will show, the energy-momentum tensor of Galileon field has complicated dependence on both scalar field quantities $\phi$, $\dot{\phi}$ etc, but also on gravitational quantities such as Riemannian tensors. This reveals that the energy-momentum tensor is not simply controlled by the property of the Galileon field itself, but also depends on the whole dynamics of the gravity-scalar field system due to the ``kinetic coupling (braiding)" between gravity and scalar field, as was also pointed in \cite{Deffayet:2010qz,Pujolas:2011he}. This fact makes the equation of state, i.e. the relation between $\rho$ and $P$ much involved.
Combining the above complexities together, one may wonder if the conservation law for the curvature perturbation still holds in Galileon models.

The present work is devoted to answer this question.
We will show the presence of a conserved \emph{non-linear}{\footnote{Here we refer to the curvature perturbation as``nonlinear" simply mean it is conserved on all orders in a perturbative analysis.}} curvature perturbation on large scales in generalized Galileon scalar field models \cite{Deffayet:2011gz}. Two different approaches are taken to prove this. The first approach is based on the energy-momentum conservation, which is the familiar approach in understanding the conservation in $k$-essence or perfect fluid models \cite{cons_gen,geocons}. We will employ the ``covariant approach" (or ``geometric approach") to cosmological perturbations \cite{geocons} (see \cite{Langlois:2010vx,Tsagas:2007yx} for reviews and \cite{cov_other} for related recent development), which can be viewed as a ``middle-way" between the usual ``coordinate-based perturbative" approach (see \cite{CPTreview} for comprehensive reviews) and the ``gradient expansion" approach \cite{grad_exp,grad_exp2,cons_gen}. Both the ``gradient expansion approach" and ``covariant approach" can produce non-perturbative results. As mentioned before, an important conclusion got from the previous analysis is that, the existence of a conserved curvature perturbation is the consequence of the conservation of energy-momentum tensor \cite{cons_gen,geocons}. Precisely, there is a conserved curvature perturbation on large scales and on all peturbative orders, as long as the ``non-adiabatic pressure" and the ``dissipative pressure" vanish, since which act as source terms in the evolution equation for the curvature perturbation. Correspondingly, the vanishing of dissipative pressure requires the fluid to be perfect, and the vanishing of non-adiabatic pressure implies the fluid is barotropic.
As we will show, this is just what happens for the  fluid corresponding the Galileon field on large scales.
To this end, we divide our proof into three parts:
    \begin{itemize}
        \item First we show the ``Galileon-fluid" is perfect on large scales, or more precisely, the dissipative pressure is vanishing. As we have mentioned, there is always a comoving frame for $k$-essencial scalar field in which its energy-momentum tensor takes the perfect fluid form. This is not possible for Galileon field. However, thanks to the higher-order derivative construction of Galileon model, both the energy-flow and the anisotropic pressure are suppressed by spatial derivatives. Consequently, the corresponding dissipative pressure is suppressed by spatial derivatives just as in $k$-essence model.
        \item Then we show the ``Galileon-fluid" is controlled by two variables $\phi$ and $\dot{\phi}$ on large scales, which is essential in our whole analysis. This is explicit in the $k$-essence model, since which depends on  $\phi$ and its kinetic term $X=-(\nabla\phi)^2/2$ and thus $\dot{\phi}$ on large scales. The situation is much more involved in Galileon model. As we have mentioned, its  energy-momentum tensor depends not only on $\phi$, $\dot{\phi}$, but also explicitly on $\ddot{\phi}$ as well as various gravitational quantities. However, after employing the Einstein equation as well as the energy-momentum conservation, we can re-express all the other quantities in terms of $\phi$ and $\dot{\phi}$. Essentially, this is because the Galileon model has only one dynamical degree of freedom, the scalar field $\phi$.
        \item Finally, we show the entropy/non-adiabatic perturbation{\footnote{Since the Galileon model involves only a single scalar field, this is often referred to as the \emph{intrinsic} entropy/non-adiabatic perturbation.}} in the Galileon model is suppressed on large scales, as what happens in $k$-essence or perfect fluid models \cite{Gordon:2000hv,Christopherson:2008ry,Langlois:2008mn}. Correspondingly, the non-adiabatic pressure vanishes and only the adiabatic mode survives on large scales, which implies the conservation of curvature perturbation.
    \end{itemize}
These three steps complete our proof.
This approach --- relying on the energy-momentum tensor and thus the fluid picture of the Galileon field (see \cite{Pujolas:2011he} for a recent investigation of the fluid picture of the Galileon term $G(X,\phi)\square\phi$) --- is physically transparent and meaningful but mathematically complicated. Thus we develop the other approach, that is to derive the evolution equation for the curvature perturbation $\zeta$ on large scales directly and to see if it possesses a constant solution. This is more familiar to the usual coordinate-based approach.

In \cite{Naruko:2011zk}, the conservation of curvature perturbation in Galileon model with $G(X,\phi)\square\phi$ term in the Lagrangian is studied, where the key assumption $\dot{\phi}=\dot{\phi}(\phi)$ is made in order to prove the conservation. We wish to emphasize is that, as we show in Sec.\ref{sec:adia} and in Appendix \ref{sec:baro_gen}, it is the suppression of entropy mode on large scales, which acts as a source term for the evolution of curvature perturbation, that ensures the conservation of curvature perturbation. Of course, as a result, the scalar field is driven to the attractor with $\dot{\phi}=\dot{\phi}(\phi)$ on large scales.

\section{Adiabaticity on large scales}{\label{sec:adia}}

\subsection{Energy-momentum conservation and evolution of curvature perturbation}

We employ the covariant approach to cosmological perturbations \cite{geocons}, and  briefly collect the relevant formulae below (see \cite{Langlois:2010vx,Tsagas:2007yx} for reviews). The covariant formalism is powerful in that, it provides an explicit and especially a covariant separation of an object into its timelike and spacelike parts, which is most convenient for our purpose for investigating the large-scale behavior of a system, where we will concentrate on the leading terms in the expansion with respect to the \emph{spatial} derivatives{\footnote{The covariant approach, which relies on the ``3+1 decomposition" --- contrary to the most popular ``coordinate-based approach", which specifies a coordinate system in which the background is spatially homogeneous, breaks the general covariance in a ``hard" manner and yields the problem of gauge artifacts ---  splits the time and space in a covariant manner without introducing any preferred coordinate system. This fact allows the covariant approach to grasp the physical transparency which is often lost in coordinate-based calculations, and to yield many interesting understandings on cosmological perturbation, including the conservation of super-Hubble curvature perturbation. However, the main obstacle in covariant approach is that, it is not easy to perform a perturbative calculation, which is needed for practical purpose.}}.

The covariant approach to cosmological perturbations is highly related to the so-called ``3+1" or covariant formalism \cite{cov_form}, which splits all covariant quantities into temporal and spatial parts by choosing a special foliation structure of spacetime. This splitting is usually realized by specifying an arbitrary unit timelike vector field $u_{a}$.
The energy-momentum tensor can be decomposed as{\footnote{Throughout this paper, symmetrization is normalized, i.e. $A_{(a}B_{b)}\equiv \frac{1}{2}(A_a B_b +A_b B_a)$. We use $a, b,\cdots$ as covariant indices, which are popular in the literatures on covariant formalism.
}}
    \begin{equation}{\label{EMT_dec}}
        T_{ab} = \rho u_{a}u_{b} +P h_{ab}  + 2 u_{(a} q_{b)} +\pi_{ab},
    \end{equation}
where $h_{ab} \equiv g_{ab}+u_a u_b$
is the spatial projection tensor which is orthogonal to the fluid velocity $u_a$, $\rho$ and $P$ are the energy density and pressure respectively, the energy flow $q_a$ and the anistropic stress $\pi_{ab}$ are both spatial tensors which satisfy $q_au^a =0$, $\pi_{ab}u^b=0$, $\pi_{ab}=\pi_{ba}$ and $g^{ab}\pi_{ab}= h^{ab}\pi_{ab}=0$.
The fluid variables can be read from (\ref{EMT_dec}) as
    \begin{eqnarray}
        \rho & \equiv & u^{a}u^{b}T_{ab},\label{rho_def}\\
        P & \equiv & \frac{1}{3}h^{ab}T_{ab},\label{P_def}\\
        q_{a} & = & -h_{a}^{b}u^{c}T_{bc},\label{qa_def}\\
        \pi_{ab} & \equiv & \left(h_{a}^{c}h_{b}^{d}-\frac{1}{3}h^{cd}h_{ab}\right)T_{cd}\label{piab_def}
        \end{eqnarray}
The spatially-projected derivative $D_a$ acting on an arbitrary tensor is defined by
    \begin{equation}
        D_{a}T_{\phantom{\cdots b\cdots}\cdots c\cdots}^{\cdots b\cdots}\equiv h_{a}^{a'}\cdots h_{b'}^{b}\cdots h_{c}^{c'}\cdots\nabla_{a'}T_{\phantom{\cdots b\cdots}\cdots c'\cdots}^{\cdots b'\cdots},
    \end{equation}
where  the operator $D_a$ is indeed a real covariant derivative, but associated with $h_{ab}$ instead of $g_{ab}$. We also introduce the decomposition of derivative of $u^a$,
    \begin{equation}{\label{Du_dec}}
        D_{b}u_{a} \equiv \nabla_{b} u_{a} + u_{b} a_{a}=\frac{1}{3}\Theta h_{ab}+\sigma_{ab} +\omega_{ab},
    \end{equation}
where $\Theta\equiv \nabla_a u^a$ is the expansion, $a_a \equiv u^b\nabla_b u_a$ is the acceleration vector, $\sigma_{ab}=\sigma_{ba}$ is the symmetric tracefree shear tensor, $\omega_{ab} = - \omega_{ba}$ is the antisymmetric vorticity tensor.
Note the Frobenius' theorem implies $\omega_{ab}=0$ if $u^a$ is hypersurface orthogonal, which is just the case we are interested in.

The projection of the conservation of energy-momentum tensor onto $u^a$, i.e. $u_a\nabla_b T^{ab} = 0$, yields a continuity equation for the  energy density and pressure:
    \begin{equation}{\label{EMT_cons}}
        \dot{\rho} +\Theta (\rho+P) = \mathcal{D},
    \end{equation}
where the dissipative term reads
    \begin{equation}{\label{curly_D_def}}
        \mathcal{D} = -q^{a}a_{a}-D_{a}q^{a}-\pi^{ab}\sigma_{ab}.
    \end{equation}
In this section, an over dot ``$\dot{\phantom{\phi}}$" denotes the Lie derivative with respect to $u^a$: $\pounds_u$, which has the natural explanation as a ``covariant" temporal derivative. When acting on scalar quantities, it is just $\dot{f} = \pounds_u f = u^a \nabla_a f$. We can define the covector{\footnote{In the covariant approach, perturbations are usually defined as spatial gradients, e.g. the scalar perturbation $\delta\phi$ in coordinate-based approach is represented by $D_a\phi$, the energy density perturbation $\delta\rho$ is represented $D_a\rho$, etc. See \cite{geocons,Langlois:2010vx,cov_form} for details.}} \cite{geocons,Langlois:2010vx}
    \begin{equation}{\label{zeta_a_def}}
        \zeta_a \equiv D_a \alpha - \frac{\dot{\alpha}}{\dot{\rho}}D_a\rho, \qquad\qquad \text{with }~ \dot{\alpha} \equiv \frac{1}{3}  \Theta,
    \end{equation}
which can be explained as the curvature perturbation. It was shown in \cite{geocons,Langlois:2010vx} that (\ref{EMT_cons}) can induce an evolution equation for $\zeta_a$,
    \begin{equation}{\label{zeta_a_eom}}
        \dot{\zeta}_a = \frac{\Theta^2}{3\dot{\rho}}\left( \Gamma_a +\Sigma_a \right),
    \end{equation}
where
    \begin{equation}{\label{Gamma_a_def}}
        \Gamma_a = D_a P - \frac{\dot{P}}{\dot{\rho}} D_a\rho,
    \end{equation}
and
    \begin{equation}{\label{Sigma_a_def}}
        \Sigma_a = -\frac{1}{\Theta}\left( D_a\mathcal{D} - \frac{\dot{\mathcal{D}}}{\dot{\rho}}D_a \rho \right) +\frac{\mathcal{D}}{\Theta^2}\left( D_a \Theta - \frac{\dot{\Theta}}{\dot{\rho}} D_a \rho \right),
    \end{equation}
are the nonlinear non-adiabatic pressure and dissipative pressure respectively{\footnote{Throughout this paper, we use the words ``dissipative pressure" or ``entropy perturbation" to indicate the effects of imperfect fluid nature of the energy-momentum tensor of the ``generalized Galileons",  which is usually used in the cosmological context. One should keep in mind that, the Galileon model is a closed and essentially conservative system, which implies there is no real physical dissipation or entropy production.}}.
Now it becomes explicit from (\ref{zeta_a_eom}) that, in order to prove the conservation of $\zeta_a$, i.e. to show it satisfies a conservation law $\dot{\zeta}_a = 0$, the source term in the right-hand-side of (\ref{zeta_a_eom}) must vanish, i.e.:
    \begin{equation}
        \Gamma_a +\Sigma_a = 0.
    \end{equation}
As we will see, this is just what happens in Galileon model on large-scales, where both the dissipative pressure $\Sigma_a$ and the non-adiabatic pressure $\Gamma_a$  vanish separately, which implies the fluid corresponding to the Galileon field becomes \emph{perfect} and \emph{barotropic} on large scales. Correspondingly, the scalar perturbation in Galileon model becomes \emph{adiabatic}.

\subsection{Galileon model with $G(X,\phi)\square\phi$}

In this section, we make a detailed investigation on the Galileon model with $G(X,\phi)\square\phi$. This term has already included the essential ingredients in our analysis, but is simple enough in order to help us to grasp the physical picture without being immersed in cumbersome mathematics. A full treatment of the generalized Galileon is presented in Sec.\ref{sec:gg}.

Consider the Galileon model
    \begin{equation}{\label{model}}
        \mathcal{L}_{\phi} = K(X,\phi) + G(X,\phi)\square \phi,
    \end{equation}
where $X\equiv -(\nabla_a\phi)^2/2$.
The corresponding energy-momentum tensor $T_{ab} \equiv -\frac{2}{\sqrt{-g}}\frac{\delta (\sqrt{-g}\mathcal{L_{\phi}})}{\delta g^{ab}}$ of the scalar field reads
    \begin{equation}{\label{EMT}}
        T_{ab} = Ag_{ab}+B\nabla_{a}\phi\nabla_{b}\phi+\nabla_{a}\phi\nabla_{b}G+\nabla_{b}\phi\nabla_{a}G,
    \end{equation}
with
    \begin{eqnarray}
        A & = & K-\nabla_{a}\phi\nabla^{a}G,\\
        B & = & K_{,X}+G_{,X}\square\phi.
    \end{eqnarray}
Here and in what follows, $K_{,X}$ denotes $\partial K/\partial X$, etc.
(\ref{EMT}) does not takes the form of the energy-momentum tensor of a perfect fluid, due to the presence of $G(X,\phi)$.
In the covariant formalism, the corresponding fluid quantities can be calculated straightforwardly:
    \begin{eqnarray}
        \rho & = & -A+\tilde{B}\dot{\phi}^{2}+2G_{,X}\dot{\phi}\dot{X},\label{rho_def}\\
        P & = & A+\frac{1}{3}\left(\tilde{B}D_{a}\phi D^{a}\phi+2G_{,X}D_{a}\phi D^{a}X\right),\label{P_def}\\
        q_{a} & = & -\tilde{B}\dot{\phi}D_{a}\phi-G_{,X}\left(\dot{X}D_{a}\phi+\dot{\phi}D_{a}X\right),\label{qa_def}\\
        \pi_{ab} & = & \tilde{B}D_{a}\phi D_{b}\phi+G_{,X}\left(D_{a}\phi D_{b}X+D_{b}\phi D_{a}X\right)\nonumber \\
         &  & -\frac{1}{3}\left(\tilde{B}D_{c}\phi D^{c}\phi+2G_{,X}D_{c}\phi D^{c}X\right)h_{ab},\label{piab_def}\end{eqnarray}
where $\tilde{B} = B+2G_{,\phi}$, $\dot{f}\equiv u^a\nabla_af$. (\ref{rho_def})-(\ref{piab_def}) can match the corresponding results derived in \cite{Pujolas:2011he} in coordinate-based approach.

It is important to note, as in the single $k$-essencial scalar field model, both the energy-flow $q_a$ and anisotropic pressure $\pi_{ab}$ involve at least one spatial derivative, and thus can be neglected on large scales. Precisely, we may write
    \begin{equation}{\label{q_pi_order}}
        q_a \sim \mathcal{O}(D),\qquad \pi_{ab} \sim \mathcal{O}(D^2),
    \end{equation}
where $\mathcal{O}(D^n)$ denotes $n$-th order in spatial derivatives.
One difference between Galileon and $k$-essence is that, in single field $k$-essence model, we are always able to choose an unique ``comoving frame" for the scalar field defined by $D_a\phi=0$, in which the fluid becomes exactly a perfect one, i.e. $q_a^{\text{(com)}} = \pi_{ab}^{\text{(com)}} = 0$. This is not the case for Galileon. As from the expression for $q_a$ and $\pi_{ab}$, in the comoving frame for the scalar field, $\pi_{ab}^{\text{(com)}}=0$ as in $k$-essence model, however the energy-flow{\footnote{This phenomenon was also found in \cite{kyy_prl} in studying the scalar perturbations of Galileon model (\ref{model}), where it was firstly pointed out that, contrary to the $k$-essence model, the  ``comoving gauge" $T^0_{i}=0$ and the ``uniform field gauge" $\delta\phi=0$ are not equivalent in Galileon model (\ref{model}).}}
    \begin{equation}
        q_a^{\text{(com)}} = -G_{,X}\dot{\phi}D_{a}X \neq 0,
    \end{equation}
due to the dependence of $G$ on $X$. As was firstly pointed out in \cite{Pujolas:2011he} and will be shown in the following sections, this is also the case in the full ``generalized Galileons". This fact prevents the energy-momentum tensor of Galileon from a perfect one.

For our purpose in this work, we concentrate on the energy density $\rho$ and the pressure$P$, whose  explicit expressions  are
    \begin{eqnarray}
        \rho = \rho(\dot{\phi},\phi,\Theta,{A}_{(\rho)}) & = & -K+\left(K_{,X}+G_{,\phi}-G_{,X}\dot{\phi}\Theta\right)\dot{\phi}^{2}+\mathcal{D}_{\left(\rho\right)},\label{rho_exp}\\
        P=P(\dot{\phi},\phi,\ddot{\phi},{A}_{(P)}) & = & K + \left(G_{,\phi}+G_{,X}\ddot{\phi}\right)\dot{\phi}^{2}+\mathcal{D}_{\left(P\right)},\label{P_exp}\end{eqnarray}
where $\mathcal{D}_{(\rho)}$ and $\mathcal{D}_{(P)}$ are terms at least second-order in spatial derivatives, i.e. $\mathcal{D}_{(\rho)}\sim\mathcal{O}(D^2)$ and $\mathcal{D}_{(P)}\sim\mathcal{O}(D^2)$, which can be safely neglected in the large-scale approximation, whose explicit expressions can be found in (\ref{D_rho})-(\ref{D_P}). In (\ref{rho_exp})-(\ref{P_exp}), the arguments ${A}_{(\rho)}$ and ${A}_{(P)}$ formally remind us the dependence of $\rho$ and $P$ on all other terms which are higher-order in spatial derivatives.
As was pointed in \cite{Deffayet:2010qz,Pujolas:2011he}, besides those terms which are higher-order in spatial derivatives, the energy density and pressure have the following functional form
    \begin{equation}
        \rho=\rho(\dot{\phi},\phi,\Theta),\qquad P=P(\dot{\phi},\phi,\ddot{\phi}).
    \end{equation}
It is useful to recall that in $k$-essence model, both the energy density and the pressure  take the form $\rho=\rho(X,\phi)$ and $P=P(X,\phi)$, which can also be written as $\rho=\rho(\dot{\phi},\phi)$ and $P=P(\dot{\phi},\phi)$ up to terms which are higher-order in spatial derivatives.
In Galileon model (\ref{model}), due to the dependence of $P$ on $\ddot{\phi}$, the pressure $P$ is determined only after solving $\ddot{\phi}$ in terms of $\dot{\phi}$, $\phi$ etc, by making use of the dynamical equations of motion. The dependence of $\rho$ on $\Theta$, however, is essential for the Galileon model, which implies the energy density depends on the dynamics of the full system, rather than solely on the property of the scalar field. For later convenience, note in(\ref{rho_exp})-(\ref{P_exp}), $\rho$ is linear in $\Theta$ and $P$ is linear in $\ddot{\phi}$ with coefficients:
    \begin{eqnarray}
        \rho_{,\Theta} &\equiv & \frac{\partial \rho(\dot{\phi},\phi,\Theta,A_{(\rho)}) }{\partial\Theta} = -G_{,X}\dot{\phi}^{3},\\
        P_{,\ddot{\phi}} &\equiv & \frac{\partial P(\dot{\phi},\phi,\ddot{\phi},A_{(P)}) }{\partial\ddot{\phi}} = G_{,X}\dot{\phi}^{2},
    \end{eqnarray}
both of which are proportional to $G_{,X}$.

From (\ref{q_pi_order}), $q_a \sim \mathcal{O}(D)$ and $\pi_{ab} \sim \mathcal{O}(D^2)$, the definition of the dissipative term (\ref{curly_D_def}) implies $\mathcal{D}\sim \mathcal{O}(D^2)$, and thus in Galileon model (\ref{model}), the dissipative pressure is third order in spatial derivative{\footnote{Readers who are familiar with the ``3+1 formalism" would recall $a_a = D_a \ln N$, where $N$ is the scalar lapse function. Thus the acceleration $a_a$ is first order in spatial derivative.}},
    \begin{equation}
        \Sigma_a \sim \mathcal{O}(D^3),
         \end{equation}
and thus can  be safely neglected on large scales, which is similar to the case in (even multi-field) $k$-essence model \cite{geocons}. Thus in order to show the conservation of $\zeta_a$ on large scales, it is essential to show $\Gamma_a \approx 0$ on large scales, which implies the fluid corresponding to Galileon (\ref{model}) becomes barotropic on large scales, equivalently, the scalar perturbation becomes adiabatic on large scales.

In the appendix \ref{sec:baro_gen}, we discuss the condition for the  barotropy of the ``2-2"-mapping system with $\rho=\rho(\dot{\phi},\phi)$ and $P=P(\dot{\phi},\phi)$, which are explicit for $k$-essence model. Here we will show that, if the universe is dominated by Galileon field, the energy density and pressure of Galileon can always be cast into the ``2-2 mapping" form, which is essential in our analysis. To this end,
we have to make use of the Einstein equation $G_{ab} = T_{ab}$. Precisely, by projecting both sides on $u^au^b$ we arrive at the energy constraint
    \begin{equation}
        u^au^b G_{ab} = \rho.
    \end{equation}
Using the decomposition (in writing this, we assume $u^a$ is hypersurface orthogonal, which is actually the case of cosmological interest)
    \begin{equation}{\label{Du}}
        D_bu_a = \frac{1}{3}\Theta h_{ab} +\sigma_{ab},
    \end{equation}
as well as the Gauss-Codacci equations (see Appendix \ref{sec:GC_rel}), we are able to rewrite the energy constraint as
    \begin{equation}{\label{en_cons}}
        \frac{1}{3}\Theta^2 + \mathcal{R} = \rho, \qquad \text{with}\qquad \mathcal{R}\equiv \frac{1}{2}\left( {}^3R - \sigma_{ab} \sigma^{ab}\right),
    \end{equation}
where ${}^3R$ is the intrinsic Ricci scalar of the spacelike hypersurface orthogonal to the 4-velocity $u_a$, which is just the Ricci scalar associated with $h_{ab}$. (\ref{en_cons}) is the covariant (and non-perturbative) version of the familiar background equation $3H^2=\rho$. In an almost FRW universe, which is the interest of cosmology, $R^{(3)}\sim \mathcal{O}(D^2)$. Similarly the shear $\sigma_{ab}$ also rapidly decreases on large scales.  Using (\ref{en_cons}) and (\ref{rho_exp}), we can solve $\Theta$ as
        \begin{eqnarray}
        \Theta & = & \Theta(\dot{\phi},\phi,{A}_{(\Theta)})\nonumber \\
        & = & -\frac{3}{2}\dot{\phi}^{3}G_{,X}\pm\sqrt{3}\sqrt{\dot{\phi}^{2}\left(G_{,\phi}+K_{,X}\right)-K
        +\frac{3}{4}G_{,X}^{2}\dot{\phi}^{6}+\mathcal{D}_{\left(\rho\right)}-\mathcal{R}},\label{Theta_sol}
        \end{eqnarray}
where ``$\pm$" denotes different possible choices of branch, ${A}_{(\Theta)}$ denotes the dependence of $\Theta$ on all terms which are higher-order in spatial derivatives.
In (\ref{Theta_sol}), it is explicit that on large scales, $\Theta$ depends only on $\dot{\phi}$, $\phi$.
Using the definition of $\rho$ together with the energy constraint (\ref{en_cons}), we can eliminate $D_a\Theta$ and $\dot{\Theta}$ and get
    \begin{eqnarray}
D_{a}\rho & = & \left(1-\frac{3\rho_{,\Theta}}{2\Theta}\right)^{-1}\left[\rho_{,\dot{\phi}}D_{a}\dot{\phi}+\rho_{,\phi}D_{a}\phi+\rho_{,A_{\left(\rho\right)}}D_{a}A_{\left(\rho\right)}-\frac{3\rho_{,\Theta}}{2\Theta}D_{a}\mathcal{R}\right],\label{Da_rho}\\
\dot{\rho} & = & \left(1-\frac{3\rho_{,\Theta}}{2\Theta}\right)^{-1}\left[\rho_{,\dot{\phi}}\ddot{\phi}+\rho_{,\phi}\dot{\phi}+\rho_{,A_{\left(\rho\right)}}\dot{A}_{\left(\rho\right)}-\frac{3\rho_{,\Theta}}{2\Theta}\dot{\mathcal{R}}\right].\label{dot_rho}\end{eqnarray}
Note even on large scales, $D_{a}\rho\neq \rho_{,\dot{\phi}}D_{a}\dot{\phi}+\rho_{,\phi}D_{a}\phi$ etc, due to $\rho_{,\Theta}\neq 0$.
As we have mentioned before, we can solve $\ddot{\phi}$ from the dynamical equation of motion for the salar field, or more conveniently through the energy-momentum conservation (\ref{EMT_cons}) together with the energy constraint (\ref{en_cons}) to get
    \begin{equation}
        \ddot{\phi} = \mathcal{A}(\dot{\phi},\phi,A_{(\ddot{\phi})}) = -\left(1+\lambda\right)^{-1}\left[\frac{\rho_{,\phi}}{\rho_{,\dot{\phi}}}\dot{\phi}+\frac{\Theta\dot{\phi}^{2}}{\rho_{,\dot{\phi}}}\left(1-\frac{3\rho_{,\Theta}}{2\Theta}\right)\left(K_{,X}+2G_{,\phi}-G_{,X}\dot{\phi}\Theta\right)+\mathcal{D}_{(\ddot{\phi})}\right],
    \end{equation}
with
    \begin{equation}{\label{lambda_def}}
        \lambda = -\frac{\Theta\rho_{,\Theta}}{\dot{\phi}\rho_{,\dot{\phi}}}\left(1-\frac{3\rho_{,\Theta}}{2\Theta}\right),
    \end{equation}
$\mathcal{D}_{(\ddot{\phi})}$ is higher-order in spatial gradient which is given in (\ref{D_ddotphi}).
Again $A_{(\ddot{\phi})}$ denotes the dependence of $\ddot{\phi}$ on terms with higher order spatial derivatives bisides $\dot{\phi}$ and $\dot{\phi}$. Thus we have
    \begin{eqnarray}
D_{a}P & = & \left(P_{,\dot{\phi}}+P_{,\ddot{\phi}}\mathcal{A}_{,\dot{\phi}}\right)D_{a}\dot{\phi}+\left(P_{,\phi}+P_{,\ddot{\phi}}\mathcal{A}_{,\phi}\right)D_{a}\phi+P_{,\ddot{\phi}}\mathcal{A}_{,A_{(\ddot{\phi})}}D_{a}A_{(\ddot{\phi})}+P_{,A_{\left(P\right)}}D_{a}A_{\left(P\right)},\label{Da_P}\\
\dot{P} & = & \left(P_{,\dot{\phi}}+P_{,\ddot{\phi}}\mathcal{A}_{,\dot{\phi}}\right)\ddot{\phi}+\left(P_{,\phi}+P_{,\ddot{\phi}}\mathcal{A}_{,\phi}\right)\dot{\phi}+P_{,\ddot{\phi}}\mathcal{A}_{,A_{(\ddot{\phi})}}\dot{A}_{(\ddot{\phi})}+P_{,A_{\left(P\right)}}\dot{A}_{\left(P\right)}.\label{dot_P}\end{eqnarray}
Although (\ref{Da_rho})-(\ref{dot_P}) look cumbersome, the meaning of whose structures is simple: besides all the terms which are higher order in spatial gradients, e.g. $A_{(\rho)}$, $\mathcal{R}$ etc, on large scales,  both $\rho$ and $P$ are only determined by $\dot{\phi}$ and $\phi$, which is essential for our following analysis.

Having shown that both the energy density and the pressure can be written as functions of $\dot{\phi}$ and $\phi$ together terms which can be neglected on large scales, we are at the point to show that $\Gamma_a \approx 0$ on large scales.
For later convenience, we introduce the ``relative entropy perturbation" between any two variables $X$ and $Y$:
    \begin{equation}{\label{S_gen_def}}
        S_a(X,Y) \equiv \frac{D_a X}{\dot{X}} - \frac{D_a Y}{\dot{Y}},
    \end{equation}
which will significantly simplifies our following discussion.
In terms of $S_a(X,Y)$, the non-adiabatic pressure is just
    \begin{equation}
        \Gamma_a = \dot{P} S_a (P,\rho).
    \end{equation}
As usual, we introduce the so-called ``comoving density perturbation"
    \begin{equation}{\label{cdp_def}}
        \epsilon_a \equiv D_a \rho - \frac{\dot{\rho}}{\dot{\phi}} D_a \phi  = \dot{\rho} S_a (\rho,\phi).
    \end{equation}
It is well-known that in the perturbation theory of perfect fluid or $k$-essence scalar field model, the comoving density perturbation is suppressed on large scales. Actually as we will show, what is suppressed is the following quantity \cite{geocons}
    \begin{equation}{\label{cdp_t_def}}
        \tilde{\epsilon}_a \equiv D_a\rho - \Theta q_a.
    \end{equation}
To this end, we have to make use of the Einstein equation. The momentum constraint is given by
    \begin{equation}
        u^b G_{bc} h^c_a = -q_a,
    \end{equation}
which can be recast into a more convenient form
    \begin{equation}{\label{m_cons}}
        D^b \sigma_{ba} - \frac{2}{3}D_a \Theta = -q_a,
    \end{equation}
through the Gauss-Codazzi relations (see Appendix \ref{sec:GC_rel}). Combining the energy constraint (\ref{en_cons}) and the momentum constraint (\ref{m_cons}) together, we get
    \begin{equation}{\label{gl_Peq}}
         D_a \mathcal{R} +\Theta D^b\sigma_{ba} = D_a\rho - \Theta q_a \equiv \tilde{\epsilon}_a,
    \end{equation}
which is the covariant generalization of the Poisson equation in linear theory. (\ref{gl_Peq}) is one of the essential equations in our analysis. In an almost FRW background, the left-hand-side of (\ref{gl_Peq}) is suppressed by spatial derivatives, which implies that $\tilde{\epsilon}_a \approx 0$ on large scales.

In perfect fluid or $k$-essence model, it was shown that $\epsilon_a$ and $\tilde{\epsilon}_a$ exactly coincide on large scales (or in linear theory) \cite{geocons}. In Galileon model, as we will show, $\epsilon_a$ and $\tilde{\epsilon}_a$ do not coincide but are proportional to each other. Explicit manipulations show that (see Appendix \ref{sec:rel_en} for detailed derivation)
    \begin{equation}{\label{te_e_rel}}
        \tilde{\epsilon}_{a}=\left(1+\lambda\right)\epsilon_{a}+\left(1+\lambda\right)\mathcal{D}_{(\epsilon_{a})}+\mathcal{D}_{(\tilde{\epsilon}_{a})},
    \end{equation}
where $\lambda$ is given in (\ref{lambda_def}),
$\mathcal{D}_{(\epsilon_{a})}$ and $\mathcal{D}_{(\tilde{\epsilon}_{a})}$ are higher-order spatial derivative terms, whose expression are given in (\ref{Dea}) and (\ref{Dtea}).
On large scales where $\mathcal{D}_{(\epsilon_{a})}\approx \mathcal{D}_{(\tilde{\epsilon}_{a})}\approx 0$, we have
    \begin{equation}{\label{te_e_rel_ls}}
        \tilde{\epsilon}_a \approx (1+\lambda) \epsilon_a.
    \end{equation}
For $k$-essence model, $\rho$ does not depend on the expansion $\Theta$ explicitly, thus $\rho_{,\Theta}=0$ and $\lambda=0$. Thus (\ref{te_e_rel_ls}) is the generalization of the relation $\tilde{\epsilon}_a \approx \epsilon_a$ in $k$-essence model to the case of Galileon (\ref{model}).
Similarly the non-adiabatic pressure can be evaluated as
    \begin{equation}{\label{Gamma_e_rel}}
        \Gamma_{a} = \frac{C\dot{\phi}}{\rho_{,\dot{\phi}}\dot{\rho}}\epsilon_{a}-\frac{C\dot{\phi}}{\rho_{,\dot{\phi}}\dot{\rho}}\mathcal{D}_{(\epsilon_{a})}+\left(1-\frac{3\rho_{,\Theta}}{2\Theta}\right)^{-1}\frac{1}{\dot{\rho}}\mathcal{S}_{a},
    \end{equation}
where
    \begin{equation}
        C= \rho_{,\phi}P_{,\dot{\phi}}-\rho_{,\dot{\phi}}P_{,\phi}+P_{,\ddot{\phi}}\left(\rho_{,\phi}\mathcal{A}_{,\dot{\phi}}-\rho_{,\dot{\phi}}\mathcal{A}_{,\phi}\right),
    \end{equation}
and $\mathcal{S}_a$ is higher-order in spatial derivatives and is given in (\ref{mS_def}).
On large scales, (\ref{Gamma_e_rel}) implies
    \begin{equation}{\label{Gamma_e_rel_ls}}
        \Gamma_a \approx \frac{C\dot{\phi}}{\rho_{,\dot{\phi}}\dot{\rho}}\epsilon_{a}.
    \end{equation}
(\ref{te_e_rel_ls}) and (\ref{Gamma_e_rel_ls}) explicitly show that on large scales $\Gamma_a$, $\epsilon_a$ and $\tilde{\epsilon}_a$ are proportional to each other. Whereas we have proved that on large scales $\tilde{\epsilon}_a$ is suppressed (see (\ref{gl_Peq})), thus we get the conclusion that, on large scales, the non-adiabatic pressure is  vanishing
    \begin{equation}
        \Gamma_a \approx 0.
    \end{equation}
This fact ensures the conservation of curvature perturbation $\zeta_a$ on large scales.

Let us try to understand the above result. The Galileon model (\ref{model}) involves only \emph{one} dynamical degree of freedom, the scalar field $\phi$. Thus \emph{on large scales}, the system involves only two perturbation modes $D_a\dot{\phi}$ and $D_a \phi$, which can be combined linearly to give one adiabatic mode and one entropy mode. The fact that the system contains only \emph{one} entropy perturbation implies that, $\Gamma_a$, $\epsilon_a$, $\tilde{\epsilon}_a$ etc, although defined differently, must be equivalent characterizations of the same entropy perturbation mode, and must be proportional to each other on large scales. What is essential is that, the entropy mode has no dynamics for itself and is  suppressed  on large scales,  only the adiabatic mode survives, as what happens in the perfect fluid or $k$-essence models.
To conclude, as in single-field $k$-essence model, the scalar perturbation in Galileon model (\ref{model}) becomes adiabatic on large scales and the corresponding curvature perturbation $\zeta_a$ is conserved.

\subsection{Generalized Galileon model}{\label{sec:gg}}

The above analysis can be applied to the generalized Galileon model proposed very recently in \cite{Deffayet:2011gz}, although the calculation becomes dramatically cumbersome due to the  increasing complexity in the structure of the energy-momentum tensor when including higher order Galileon terms.

The full Lagrangian of the generalized Galileon is given by \cite{Deffayet:2011gz}
    \begin{equation}{\label{action}}
        S = \int d^4x \sqrt{-g} \left( \frac{1}{2}R + \sum_{n=0}^{3}\mathcal{L}^{(n)} \right),
    \end{equation}
with
    \begin{eqnarray}
        \mathcal{L}^{(0)} & = & K^{\left(0\right)}\left(X,\phi\right),\\
        \mathcal{L}^{(1)} & = & K^{\left(1\right)}\left(X,\phi\right)\Box\phi,\\
        \mathcal{L}^{(2)} & = & K^{\left(2\right)}\left(X,\phi\right)\left[\left(\Box\phi\right)^{2}-\left(\nabla_{a}\nabla_{b}\phi\right)^{2}\right]
        +R\, Q^{\left(2\right)}\left(X,\phi\right) ,\\
        \mathcal{L}^{(3)} & = & K^{\left(3\right)}\left(X,\phi\right)\left[\left(\Box\phi\right)^{3}-3\Box\phi\left(\nabla_{a}\nabla_{b}\phi\right)^{2}
        +2\left(\nabla_{a}\nabla_{b}\phi\right)^{3}\right] \nonumber\\
        &&-6G_{ab}\nabla^{a}\nabla^{b}\phi\,
        Q^{\left(3\right)}\left(X,\phi\right) ,
        \end{eqnarray}
where $X=-(\nabla\phi)^2/2$
and
    \begin{equation}
        Q^{\left(2,3\right)}\left(X,\phi\right) \equiv \int_{0}^{X}dY\,K^{\left(2,3\right)}\left(Y,\phi\right),
    \end{equation}
that is $Q^{(n)}_{,X}\equiv K^{(n)}$ and $K^{(n)}(X,\phi)$ etc are arbitrary functions of $X$ and $\phi$, $R$ and $G_{ab}$ are Ricci scalar and Einstein tensor respectively.

The energy-momentum tensor of the generalized Galileon model (\ref{action}) is given by
    \begin{equation}
        T_{ab} = \sum_{n=0}^{3} T^{(n)}_{ab},\qquad\qquad T^{(n)}_{ab} \equiv  -\frac{2}{\sqrt{-g}}\frac{\delta (\sqrt{-g}\mathcal{L}^{(n)})}{\delta g^{ab}}
    \end{equation}
where a superscript ``${}^{(n)}$" denotes the contribution from $\mathcal{L}^{(n)}$. We have
    \begin{eqnarray}
        T_{ab}^{(0)} & = & K^{\left(0\right)}g_{ab}+K_{,X}^{\left(0\right)}\nabla_{a}\phi\nabla_{b}\phi,\label{EMT0}\\
        T_{ab}^{(1)} & = & -\left(K_{,X}^{(1)}\nabla^{c}\phi\nabla_{c}X-2XK_{,\phi}^{(1)}\right)g_{ab} \nonumber\\
        &&+\left(K_{,X}^{\left(1\right)}\square\phi+2K_{,\phi}^{(1)}\right)\nabla_{a}\phi\nabla_{b}\phi+2K_{,X}^{(1)}\nabla_{(a}\phi\nabla_{b)}X,\label{EMT1}
        \end{eqnarray}
where $T^{(0)}_{ab}+T^{(1)}_{ab}$ is just (\ref{EMT}) with $K^{(0)}=K$ and $K^{(1)}=G$. Contributions from $\mathcal{L}^{(2)}$ and $\mathcal{L}^{(3)}$ are much involved, which are given by
    \begin{eqnarray}
        T_{ab}^{(2)} & = & g_{ab}\Big\{ RQ^{\left(2\right)}-K^{(2)}\left(\left(\square\phi\right)^{2}-\left(\nabla_{c}\nabla_{d}\phi\right)^{2}\right)+4XQ_{,\phi\phi}^{(2)}-2K_{,X}^{(2)}\nabla_{c}X\nabla^{c}X\nonumber \\
         &  & -4K_{,\phi}^{(2)}\nabla_{c}\phi\nabla^{c}X-2\square\phi\left(K_{,X}^{(2)}\nabla_{c}\phi\nabla^{c}X-2XK_{,\phi}^{(2)}+Q_{,\phi}^{(2)}\right)+2K^{(2)}R_{cd}\nabla^{c}\phi\nabla^{d}\phi\Big\}\nonumber \\
         &  & +\left[K^{(2)}R+4\square\phi K_{,\phi}^{(2)}+2Q_{,\phi\phi}^{(2)}+K_{,X}^{(2)}\left(\left(\square\phi\right)^{2}-\left(\nabla_{c}\nabla_{d}\phi\right)^{2}\right)\right]\nabla_{a}\phi\nabla_{b}\phi\nonumber \\
         &  & +4\left(\square\phi K_{,X}^{(2)}+2K_{,\phi}^{(2)}\right)\nabla_{(a}\phi\nabla_{b)}X+2K_{,X}^{(2)}\left(\nabla_{a}X\nabla_{b}X-2\nabla_{c}X\nabla^{c}\nabla_{(a}\phi\nabla_{b)}\phi\right)\nonumber \\
         &  & +2\left(K^{(2)}\square\phi+K_{,X}^{(2)}\nabla_{c}\phi\nabla^{c}X-2XK_{,\phi}^{(2)}+Q_{,\phi}^{(2)}\right)\nabla_{a}\nabla_{b}\phi\nonumber \\
         &  & -2K^{(2)}\left(\nabla^{c}\nabla_{a}\phi\nabla_{b}\nabla_{c}\phi+2\nabla_{(a}\phi R_{b)c}\nabla^{c}\phi+R_{cadb}\nabla^{c}\phi\nabla^{d}\phi\right)-2Q^{(2)}R_{ab},\label{EMT2}
        \end{eqnarray}
and $T^{(3)}_{ab}$ is given in Appendix \ref{sec:EMT3}.
As $T^{(1)}_{ab}$ depends on $\Theta$, $T^{(2)}_{ab}$ and $T^{(3)}_{ab}$ have explicit dependence on gravitational objects such as $R$, $R_{ab}$ and $R_{abcd}$ etc.

Now we concentrate on the large scale behavior of the energy-momentum tensor.
To this end, one would find the following two expressions be useful
    \begin{eqnarray}
        \nabla_{a}\phi & = & -u_{a}\dot{\phi}+D_{a}\phi,\label{nabla_phi_dec}\\
        \nabla_{a}\nabla_{b}\phi & = & u_{a}u_{b}\left(\ddot{\phi}-a^{c}D_{c}\phi \right)-2u_{(a}\left(D_{b)}\dot{\phi}+D^{c}u_{b)}D_{c}\phi\right)-D_{b}u_{a}\dot{\phi}+D_{a}D_{b}\phi,\label{nabla2_phi_dec}
        \end{eqnarray}
where $D_bu_a$ is given by (\ref{Du}). By keeping only the leading terms in spatial derivatives, we can write
    \begin{eqnarray}
        \nabla_{a}\phi & = & -u_{a}\dot{\phi}+\mathcal{O}(D), {\label{nabla_phi_ls}}\\
        \nabla_{a}\nabla_{b}\phi & = & u_{a}u_{b}\ddot{\phi}-\frac{1}{3}\Theta\dot{\phi}h_{ab}+\mathcal{O}(D),{\label{nabla2_phi_ls}}
        \end{eqnarray}
where again, $\mathcal{O}(D)$ denotes higher-order spatial derivative terms which will be neglected below.
For later convenience, note
    \begin{equation}{\label{X_nablaX_ls}}
        X = \frac{1}{2}\dot{\phi}^2 +\mathcal{O}(D^2),\qquad\qquad \nabla_a X = -u_a \dot{\phi} \ddot{\phi} +\mathcal{O}(D),
    \end{equation}
which can be verified easily. At this point, it is useful to note, besides $u^a$ and $h_{ab}$, $\nabla_a \phi$ and $X$ depend on $\dot{\phi}$, while $\nabla_a \nabla_b \phi$ and $\nabla_a X$ depend on $\dot{\phi}$, $\ddot{\phi}$ and $\Theta$ on large scales.

Plugging (\ref{nabla_phi_ls})-(\ref{nabla2_phi_ls}) into (\ref{EMT0})-(\ref{EMT1}), we have
    \begin{eqnarray}
        T^{(0)}_{ab} & \approx & \left(K_{,X}^{\left(0\right)}\dot{\phi}^{2}-K^{\left(0\right)}\right)u_{a}u_{b}+K^{\left(0\right)}h_{ab},{\label{EMT0_ls}}\\
        T^{(1)}_{ab} & \approx & \left(K_{,\phi}^{(1)}-K_{,X}^{\left(1\right)}\Theta\dot{\phi}\right)\dot{\phi}^{2}u_{a}u_{b}+\left(K_{,X}^{(1)}\ddot{\phi}+K_{,\phi}^{(1)}\right)\dot{\phi}^{2}h_{ab},{\label{EMT1_ls}}
    \end{eqnarray}
from which the corresponding large-scale expressions for the  energy density and pressure $\rho^{(0)}$, $\rho^{(1)}$, $P^{(0)}$ and $P^{(1)}$ can be read easily.
From (\ref{EMT0_ls})-(\ref{EMT1_ls}), it is explicit that $T^{(0)}_{ab}$ and $T^{(1)}_{ab}$ become that of a \emph{perfect} fluid separately (and automatically) in the large-scale approximation.
Actually, (\ref{nabla_phi_ls}) and thus (\ref{EMT0_ls}) are \emph{exact} in the comoving frame of $\phi$ where $D_a\phi=0$, which yields the well-known conclusion that $k$-essence scalar field in its own comoving frame corresponds to a perfect fluid.

Similar procedures can be applied to $T^{(2)}_{ab}$ and $T^{(3)}_{ab}$. In the following, we perform a detailed analysis on $T^{(2)}_{ab}$. The large-scale expression for $T^{(2)}_{ab}$ is
    \begin{eqnarray}
        T_{ab}^{(2)} & \approx & u_{a}u_{b}\left[-\left(\frac{4}{3}\Theta^{2}+2\dot{\Theta}\right)Q^{\left(2\right)}+4K^{(2)}\left(\frac{2}{3}\Theta^{2}+\dot{\Theta}\right)\dot{\phi}^{2}+K_{,X}^{(2)}\frac{2}{3}\Theta^{2}\dot{\phi}^{4}-2K_{,\phi}^{(2)}\Theta\dot{\phi}^{3}-2\Theta\dot{\phi}Q_{,\phi}^{(2)}\right]\nonumber \\
         &  & +h_{ab}\left[\left(\frac{4}{3}\Theta^{2}+2\dot{\Theta}\right)Q^{\left(2\right)}+K^{(2)}\left(-\frac{8}{9}\Theta^{2}\dot{\phi}^{2}-\frac{4}{3}\Theta\dot{\phi}\ddot{\phi}-2\dot{\Theta}\dot{\phi}^{2}\right)-\frac{4}{3}K_{,X}^{(2)}\ddot{\phi}\dot{\phi}^{3}\Theta\right.\nonumber \\
         &  & \qquad\left.+2\dot{\phi}^{2}Q_{,\phi\phi}^{(2)}+2K_{,\phi}^{(2)}\left(\ddot{\phi}-\frac{2}{3}\Theta\dot{\phi}\right)\dot{\phi}^{2}+2\left(\ddot{\phi}+\frac{2}{3}\Theta\dot{\phi}\right)Q_{,\phi}^{(2)}\right]\nonumber \\
         &  & \qquad-2K^{(2)}\left(2u_{(a}R_{b)c}u^{c}\dot{\phi}^{2}+R_{cadb}u^{c}u^{d}\dot{\phi}^{2}\right)-2Q^{(2)}R_{ab}.\label{EMT2_ls}
        \end{eqnarray}
where we have used the decomposition of $R$, $R_{cd}u^{c}u^{d}$ on large scales, see Appendix \ref{sec:GC_rel} for details.
$T^{(2)}_{ab}$ have almost taken the form of a perfect fluid, except the last line. Instead of trying to show the last line in (\ref{EMT2_ls}) can also be written in a perfect form,
it is easier, however, to evaluate the corresponding energy-flow $q^{(2)}_a$ and the anisotropic stress $\pi^{(2)}_{ab}$ directly and to see if they are indeed vanishing on large scales. Only the last line in (\ref{EMT2_ls}) contributes to $q^{(2)}_a$ and $\pi^{(2)}_{ab}$, thus we have
    \begin{eqnarray}
        q_{a}^{(2)} &=& -2\left(K^{(2)}\dot{\phi}^{2}-Q^{(2)}\right)h_{a}^{b}u^{c}R_{bc}+\mathcal{O}(D),{\label{qa_2}}\\
        \pi^{(2)}_{ab} &= & -2\left(h_{a}^{c}h_{b}^{d}-\frac{1}{3}h^{cd}h_{ab}\right)\left(K^{(2)}\dot{\phi}^{2}R_{ecfd}u^{e}u^{f}+Q^{(2)}R_{cd}\right) +\mathcal{O}(D^2) .
    \end{eqnarray}
Contrary to $T^{(0)}_{ab}$ and $T^{(1)}_{ab}$, at this point it is not explicit if $q_a \approx \pi_{ab} \approx 0$ and thus $T^{(2)}_{ab}$ takes the form of a perfect fluid under the large-scale approximation. Now we need to employ the Gauss-Codazzi relations. (\ref{codazzi_con}) yields
    \begin{equation}
        h_{a}^{b}u^{c}R_{bc} = D_{b}\sigma_{a}^{b} -  \frac{2}{3}D_{a}\Theta \sim \mathcal{O}(D),
    \end{equation}
which implies
    \begin{equation}
        q^{(2)}_a \sim \mathcal{O}(D),
        \end{equation}
and thus vanishes in the large-scale approximation.
It is interesting to note, a special case is when the coefficient in (\ref{qa_2}) vanishes, i.e.
    \[
        K^{(2)}\dot{\phi}^{2}-Q^{(2)} =0.
    \]
In this case $q_a^{(2)}$ vanishes automatically without using the Codazzi relation. This is satisfied when $K^{(2)} \propto X$, which corresponds the original Galileon model studied in \cite{Nicolis:2008in}.
Similarly analysis can be applied to $\pi^{(2)}_{ab}$. To this end, we need to use the Ricci equation (\ref{Ricci_eq}) as well as (\ref{Ricci_s_dec}), (\ref{uuRicciT}) and (\ref{hhRicciT}), which yield
    \begin{eqnarray}
        \left(h_{a}^{c}h_{b}^{d}-\frac{1}{3}h^{cd}h_{ab}\right)R_{ecfd}u^{e}u^{f} & = & -\frac{1}{9}\left(\Theta^{2}+3\dot{\Theta}\right)h_{ab}+\frac{1}{3}h_{ab}\left(\frac{\Theta^{2}}{3}+\dot{\Theta}\right)+\mathcal{O}(D^{2}),\\
        \left(h_{a}^{c}h_{b}^{d}-\frac{1}{3}h^{cd}h_{ab}\right)R_{cd} & = & \frac{1}{3}\left(\Theta^{2}+\dot{\Theta}\right)h_{ab}-\frac{1}{3}\left(\frac{4}{3}\Theta^{2}+2\dot{\Theta}-\frac{\Theta^{2}}{3}-\dot{\Theta}\right)h_{ab}+\mathcal{O}(D^{2}),
        \end{eqnarray}
which cancel exactly and leave us terms of order $\mathcal{O}(D^2)$. Thus, we can conclude that
    \begin{equation}
        \pi^{(2)}_{ab} \sim \mathcal{O}(D^2),
    \end{equation}
as what happens for $\pi^{(1)}_{ab}$. Similar procedures can be applied to $T^{(3)}_{ab}$ and yield the same conclusion.
To conclude, the energy-flow $q_a$  and dissipative pressure $\pi_{ab}$ in the generalized Galileon model (\ref{action}) are higher-order terms in spatial derivatives and thus can be safely neglected in the large-scale approximation, which implies the fluid corresponding to the Galileon model (\ref{action}) becomes a \emph{perfect} one on large scales.
One may notice that up to now, the Einstein equation has never been used. Thus the above conclusion is generic, which is irrelevant to the underlying theory of gravitation.

Having shown both $T^{(2)}_{ab}$ and $T^{(3)}_{ab}$ become the form of a perfect fluid on large scales, which implies the corresponding contributions to the dissipative pressure vanish $\Sigma_a\approx 0$, the next step is to show the fluids described by $T^{(2)}_{ab}$ and $T^{(3)}_{ab}$  are actually \emph{barotropic}, and thus the corresponding non-adiabatic pressure also vanishes $\Gamma_a\approx 0$.
From (\ref{EMT2_ls}), the energy density and pressure from $T^{(2)}_{ab}$ can be evaluated straightforwardly{\footnote{Actually, these large-scale expressions for $\rho$ and $P$ can also be derived  through simply the background equations of motion by identifying $\Theta=3H$.}}
    \begin{equation}
        \rho^{(2)}\approx-\frac{2}{3}Q^{(2)}\Theta^{2}+\frac{4}{3}K^{(2)}\Theta^{2}\dot{\phi}^{2}+\frac{2}{3}K_{,X}^{(2)}\Theta^{2}\dot{\phi}^{4}-2Q_{,\phi}^{(2)}\Theta\dot{\phi}-2K_{,\phi}^{(2)}\Theta\dot{\phi}^{3},\label{rho2_ls}
        \end{equation}
and
    \begin{eqnarray}
        P^{(2)} & \approx & \frac{2}{3}Q^{(2)}\left(\Theta^{2}+2\dot{\Theta}\right)-\frac{2}{3}K^{(2)}\dot{\phi}\left(\Theta^{2}\dot{\phi}+2\dot{\Theta}\dot{\phi}+2\Theta\ddot{\phi}\right)-\frac{4}{3}K_{,X}^{(2)}\Theta\dot{\phi}^{3}\ddot{\phi}\nonumber \\
         &  & +2Q_{,\phi\phi}^{(2)}\dot{\phi}^{2}+2Q_{,\phi}^{(2)}\left(\ddot{\phi}+\frac{2}{3}\Theta\dot{\phi}\right)+2K_{,\phi}^{(2)}\dot{\phi}^{2}\left(\ddot{\phi}-\frac{2}{3}\Theta\dot{\phi}\right),\label{P2_ls}
        \end{eqnarray}
where again we used various relations in Appendix \ref{sec:GC_rel}.  Similarly, for $T^{(3)}_{ab}$,
    \begin{equation}
        \rho^{(3)}\approx-\frac{10}{9}K^{(3)}\Theta^{3}\dot{\phi}^{3}-\frac{2}{9}K_{,X}^{(3)}\Theta^{3}\dot{\phi}^{5}+6Q_{,\phi}^{(3)}\Theta^{2}\dot{\phi}^{2}+2K_{,\phi}^{(3)}\Theta^{2}\dot{\phi}^{4},\label{rho3_ls}
        \end{equation}
and
    \begin{eqnarray}
        P^{(3)} & \approx & \frac{2}{9}K^{(3)}\Theta\dot{\phi}^{2}\left(2\Theta^{2}\dot{\phi}+6\dot{\Theta}\dot{\phi}+9\Theta\ddot{\phi}\right)+\frac{2}{3}K_{,X}^{(3)}\Theta^{2}\dot{\phi}^{4}\ddot{\phi}\nonumber \\
         &  & -4Q_{,\phi\phi}^{(3)}\Theta\dot{\phi}^{3}-2Q_{,\phi}^{(3)}\dot{\phi}\left( \Theta^{2}\dot{\phi}+2\dot{\Theta}\dot{\phi}+4\Theta\ddot{\phi}\right)+\frac{2}{3}K_{,\phi}^{(3)}\Theta\dot{\phi}^{3}\left(\Theta\dot{\phi}-6\ddot{\phi}\right).\label{P3_ls}
        \end{eqnarray}

Now we concentrate on the functional structure of $\rho$ and $P$ in generalized Galileon model (\ref{action}).
On large scales, note $K^{(n)}$, $Q^{(n)}$ etc are functions of $\phi$ and $X\approx\dot{\phi}^2/2$ and thus (see (\ref{EMT0_ls})-(\ref{EMT1_ls}), (\ref{rho2_ls})-(\ref{P3_ls})) the energy density $\rho$ and the pressure $P$  take the following general functional form
    \begin{eqnarray}
         \rho &=& \rho_1 + \Theta\rho_2+ \Theta^2 {\rho}_3 +\mathcal{O}(D),\\
         P &=& P_1 + \Theta P_2+ \Theta^2 P_3 + \ddot{\phi} (P_4 +\Theta {P}_5 + \Theta^2 P_6) + \dot{\Theta}(P_7 +\Theta P_8) +\mathcal{O}(D),
    \end{eqnarray}
where $\rho_i$ and $P_i$ are functions of $\phi$ and $\dot{\phi}$, whose explicit expressions can be read from (\ref{EMT0_ls})-(\ref{EMT1_ls}), (\ref{rho2_ls})-(\ref{P3_ls}). Let us focus on the dependence on $\Theta$, $\dot{\Theta}$ and $\ddot{\phi}$ of $\rho$ and $P$.
On one hand, $\rho$ is simply a  binomial of $\Theta$, which does not depend on $\dot{\Theta}$ and $\ddot{\phi}$.
On the other hand, $\dot{\Theta}$ and $\ddot{\phi}$ enter $P$ only linearly.
The point is, this type of functional forms of $\rho$ and $P$ allows us to solve $\Theta$, $\dot{\Theta}$ and $\ddot{\phi}$ in terms of $\phi$ and $\dot{\phi}$. To this end,  the Einstein equation and the energy-momentum constraint (which is equivalent to the evolution equation for the Galileon field) have to be employed. Precisely, we get two equations from the Einstein equation:
    \begin{eqnarray}
        \rho & = &u^au^b G_{ab} = \frac{1}{3}\Theta^2 + \frac{1}{2}\left( R^{(3)} - \sigma_{ab} \sigma^{ab}\right) = \frac{1}{3}\Theta^2 +\mathcal{O}(D), \\
        3P & = & h^{ab}G_{ab} =-\Theta^{2}-2\dot{\Theta}-\frac{1}{2}R^{(3)}-\frac{3}{2}\sigma_{ab}\sigma^{ab}+2a_{a}a^{a}+2D_{a}a^{a}\approx -\Theta^{2}-2\dot{\Theta}+\mathcal{O}(D^2),
    \end{eqnarray}
where we used relations in Appendix \ref{sec:GC_rel}.
Alternatively, one may employ the energy-momentum constraint (\ref{EMT_cons}). In any case, we can solve $\Theta$ and $\ddot{\phi}$ in terms of $\phi$ and $\dot{\phi}$ and thus finally arrive at:
    \begin{equation}
        \rho=\rho(\phi,\dot{\phi}),\qquad\qquad P=P(\phi,\dot{\phi}),
    \end{equation}
just as what we did for $G(X,\phi)\square\phi$ model. It is this fact that both the energy density and pressure can be cast into a ``2-2" mapping that ensures the barotropy on large scales, as is discussed in Appendix \ref{sec:baro_gen}. The key point is that, as what happens  for $K(X,\phi) +G(X,\phi)\square\phi$ model (see (\ref{gl_Peq})), the non-adiabatic/entropy perturbation mode is always suppressed on large scales, leaving us only the adiabatic mode. Actually as emphasized in Appendix \ref{sec:baro_gen} and \ref{sec:supp_en}, this is a general conclusion as long as $\rho$ and $P$ are determined by \emph{two} variables $\phi$ and $\dot{\phi}$, which is irrelevant to the details of the model.

To end this section, we would like to make a general comment in order to understand why we can always get $\rho=\rho(\phi,\dot{\phi})$ and $P=P(\phi,\dot{\phi})$, which is essential in our analysis.
The Galileon action (\ref{action}) and the corresponding energy-momentum tensor $T_{ab}$  are generally constituted by the following ingredients:
    \[
        \phi,\qquad \nabla_a\phi, \qquad \nabla_a\nabla_b \phi,\qquad R_{abcd},
    \]
as well as the metric $g_{ab}$. In the scalar field sector, from (\ref{nabla_phi_ls}) and (\ref{nabla2_phi_ls}), on large scales $\nabla_a \phi$ is proportional to $\dot{\phi}$ and $\nabla_a \nabla_b\phi$ is function of $\dot{\phi}$, $\ddot{\phi}$ and $\Theta$. In the gravity sector, (\ref{Gauss_con})-(\ref{hhRicciT}), in all contractions of Riemann tensor with $\nabla_a\phi$, $\nabla_a\nabla_b\phi$ and $g_{ab}$ (and thus with $h_{ab}$ and $u^a$) which will contribute to the energy-momentum tensor, only terms proportional to $\Theta^2$ and $\dot{\Theta}$ survive on large scales. Thus, the system is apparently controlled by
    \[
        \phi, \quad\dot{\phi}, \quad \ddot{\phi}, \quad \Theta, \quad \dot{\Theta},
        \]
on large scales.
However, the Galileon model (\ref{action}) is constructed in such a very special manner that the resulting equations of motion are second order, both for the Galileon field and the metric. Two implications are in order:
    \begin{itemize}
        \item Mathematically, repeated coordinate indices only appear once, which implies that the second order temporal derivative $\ddot{\phi}$ enters the theory only \emph{linearly}{\footnote{It is also this fact makes the Galileon model has a well-defined Cauchy problem.}}, which can be verified explicitly, as
            \begin{eqnarray}
                \square\phi & \approx & -\ddot{\phi}-\Theta\dot{\phi},\\
                \left(\square\phi\right)^{2}-\left(\nabla_{a}\nabla_{b}\phi\right)^{2} &\approx & 2\Theta\dot{\phi}\ddot{\phi}+\frac{2}{3}\Theta^{2}\dot{\phi}^{2},\\
                \left(\Box\phi\right)^{3}-3\Box\phi\left(\nabla_{a}\nabla_{b}\phi\right)^{2}+2\left(\nabla_{a}\nabla_{b}\phi\right)^{3} &\approx & -2\Theta^{2}\dot{\phi}^{2}\ddot{\phi}-\frac{2}{9}\Theta^{3}\dot{\phi}^{3},
            \end{eqnarray}
        all of which are linear in $\ddot{\phi}$. This is also what happens in the energy-momentum tensor as well as the equations of motion.
    \item On the other hand, the second-order-preserving construction prevents Galileon from any extra dynamical degree of freedom, which means $\Theta$ is not a dynamical degree of freedom and can always be solved in terms of the real dynamical degrees of freedom. Mathematically, this fact implies in the equations of motion (and thus in the energy-momentum tensor and Einstein tensor), there is no temporal derivatives of $\Theta$ higher than the first order; and $\dot{\Theta}$ only appear \emph{linearly}, as we have seen.
        \end{itemize}
These two facts allow us to solve both $\Theta$ (and thus $\dot{\Theta}$) and $\ddot{\phi}$ in terms of $\phi$ and $\dot{\phi}$, at least in principle. Physically, this is simply because the real dynamical degrees of freedom in Galileon model (\ref{action}) on large scales are $\phi$ and $\dot{\phi}$.

To conclude, the fluid corresponding to the Galileon model (\ref{action}) becomes both perfect and barotropic on large scales. This fact ensures the adiabaticity of the scalar perturbations in our model on large scales and thus the conservation of curvature perturbation.

\section{Shift-symmetry on large scales}

In the above section, we have shown the existence of a conserved quantity which has the physical meaning as the curvature perturbation, using the conservation of energy-momentum tensor of the scalar field. The proof is physically transparent and meaningful but is mathematically involved. Thus it would be interesting to see the behavior of the curvature perturbation on large scales by deriving its own evolution equation, and to verify if it indeed possesses a constant solution.

One may recall that in $k$-essence models, the quadratic order action for the curvature perturbation $\zeta$ (defined by $\delta g_{ij} = a^2(1+2\zeta)\delta_{ij}$) is $S_{(2)} = \int d\eta d^3 x a^2\frac{\epsilon}{c_s^2}\left[ \zeta'^2-c_s^2(\partial_i\zeta)^2\right]$.  On large-scales (neglecting spatial derivatives) the action reduces to an action for one-dimensional system $S_{(2)} \propto \int d\eta  a^2\frac{\epsilon}{c_s^2}\zeta'^2_{\text{L}}$ for $\zeta_{\text{L}}=\zeta_{\text{L}}(\eta)$, where $\zeta_{\text{L}}(\eta)$ is the long wavelength component of the curvature perturbation, which can be viewed as the averaged curvature perturbation of a given local Hubble volume. The corresponding equation of motion for $\zeta_{\text{L}}$ is $\left(  a^2\frac{\epsilon}{c_s^2}\zeta'_{\text{L}}\right)'=0$, which has a dominant solution $\zeta_{\text{L}} = \text{const.}$ during inflation. This indicates the conservation of $\zeta_{\text{L}}$, i.e. the conservation of $\zeta$ on large-scales at \emph{linear} order. At this point, it is interesting to note that the absence of ``mass term" $\zeta^2$ in its action ensures the conservation of $\zeta$. Actually, due to the absence of ``mass term", the large-scale Lagrangian $a^2\frac{\epsilon}{c_s^2}\zeta'^2$ has the ``shift symmetry" $\zeta\rightarrow \zeta+\text{const}$, which is responsible to the conservation of $\zeta$.

Following this logic, now our task is to show $\zeta_{\text{L}} = \text{const.}$ is actually a solution (dominant solution in an expanding background) of the Galileon models on \emph{non-linear} orders. As we will see, the shift symmetry $\zeta\rightarrow \zeta+\text{const}$ is preserved on fully non-linear orders, which is essential for the conservation of $\zeta$.

We work in uniform field gauge with $\delta\phi=0$, and since we will study the perturbation on the large scales, we take the metric perturbation as
    \begin{equation}
        ds^{2}=a^{2}\left(-e^{2\alpha}d\eta^{2}+e^{2\zeta}\delta_{ij}dx^{i}dx^{j}\right),
    \end{equation}
where $a=a(\eta)$ is the scale factor and $\zeta$ will be identified as the curvature perturbation. The vector modes have no dynamics for itself (do not propagate), which decay on large scales in scalar field models. Tensor modes will couple to scalar mode on nonlinear orders. However, these couplings arise only through the spatial derivatives of scalar modes, e.g. $\partial_i\zeta$, which can be safely neglected for our purpose. In the following, we neglect spatial dependence of $\alpha$ and $\zeta$, which means we focus on their long wavelength components $\alpha_{\text{L}}$ and $\zeta_{\text{L}}$, which are spatially uniform in a local Hubble volume. In the following we neglect the subscript ``$\text{L}$" for short.

As a warm-up excise, let us consider the simplest case with $\mathcal{L}_{\phi}=X$.
The background equations are $6\mathcal{H}^{2}=\phi'^{2}$ and $\mathcal{H}'+2\mathcal{H}^2=0$, where a prime denotes derivative with respect to comoving time $\eta$.
In this case, the large-scale Lagrangian{\footnote{In this section, Lagrangian $\mathcal{L}$ includes $\sqrt{-g}$.}} is
    \begin{equation}{\label{L_X}}
        \mathcal{L} \approx-3a^{2}e^{-\alpha+3\zeta}\left(\zeta'^{2}+2\mathcal{H}\zeta'\right).
    \end{equation}
Varying (\ref{L_X}) with respect to $\alpha$ yields an equation for $\zeta$:
    \begin{equation}{\label{X_zeta_eom}}
        e^{3\zeta} \zeta' \left(2\mathcal{H}+\zeta'\right)=0,
    \end{equation}
which implies $\zeta'=0$ and thus the conservation of $\zeta$. At this point, we wish to emphasize that (\ref{X_zeta_eom}) is a fully non-perturbative result, and thus $\zeta=\text{const.}$ is a non-perturbative solution of the system on large-scales.
For completeness, it is interesting to solve the constraint $\alpha$ in terms of $\zeta$.
Varying (\ref{L_X}) with respect to $\zeta$ yields an equation:
    \begin{equation}
        \alpha'\left(\zeta'+\mathcal{H}\right)-2\mathcal{H}\zeta'-\zeta''-\frac{3}{2}\zeta'^{2}=0,
    \end{equation}
from which we can solve
    \begin{equation}{\label{X_alpha_sol}}
        \alpha = \ln\left( 1+\frac{\zeta'}{\mathcal{H}} \right) + \frac{3}{2}\int d\eta\frac{\zeta'^{2}}{\zeta'+\mathcal{H}}.
    \end{equation}
Expanding (\ref{X_alpha_sol}) perturbatively gives the familiar results \cite{Maldacena:2002vr}: $\alpha = \frac{\zeta'}{\mathcal{H}}+\cdots$.

\subsection{Galileon model with $G(X,\phi)\square\phi$}

Now we turn to the Galileon model (\ref{model}). After some manipulations, we arrive at a fully \emph{non-perturbative} action for the curvature perturbation $\zeta$ and $\alpha$ on the large scales:
    \begin{equation}{\label{S_zeta}}
        S = \int d\eta a^{2}e^{-\alpha+3\zeta}\Big[-3\left(\mathcal{H}+\zeta'\right)^{2} +a^{2}e^{2\alpha}P\left(X,\phi\right)+G'\left(X,\phi\right)\phi'\Big],
        \end{equation}
where on large scales $X\approx\frac{1}{2}\frac{e^{-2\alpha}}{a^{2}}\phi'^{2}$.
We emphasize that (\ref{S_zeta}) is \emph{exact} in the limit of large scales, by neglecting spatial gradients. No other approximation is made in deriving (\ref{S_zeta}).
Mathematically, as we have already emphasized, the large scale action (\ref{S_zeta}) only involves $\zeta$ through an overall exponential factor. Thus under the constant shift of $\zeta$:
    \begin{equation}
        \zeta \rightarrow \zeta+ c,
    \end{equation}
where $c$ is an arbitrary non-vanishing constant, the Lagrangian changes as
    \begin{equation}
        \mathcal{L}\rightarrow e^{3c} \mathcal{L},
    \end{equation}
which has no dynamical meaning and implies the system is actually invariant under the constant shift of $\zeta$.
This is just the case of $k$-essence model we mentioned before, while the presence of Galileon-like term $G(X,\phi)\square\phi$ does not change this property. To conclude, if $\zeta=0$ is a solution of the system, $\zeta=\text{const}\neq 0$ must also be a solution.

Although the above argument has already completed our proof, one may convince him/herself by deriving explicitly the evolution equations. Varying (\ref{S_zeta}) with respect to $\alpha$ and $\zeta$ yields
    \begin{equation}
        0=3\left(\mathcal{H}+\zeta'\right)^{2}+\frac{3G_{,X}}{a^{2}e^{2\alpha}}\phi'^{3}\left(\mathcal{H}+\zeta'\right)+a^{2}e^{2\alpha}P-\left(P_{,X}+G_{,\phi}\right)\phi'^{2}, {\label{alpha_eom}}
    \end{equation}
and
    \begin{eqnarray}
        0 & = &\left(\mathcal{H}+\zeta'\right)\left(\mathcal{H}+3\zeta'-2\alpha'\right)+a^{2}e^{2\alpha}P+2\left(\mathcal{H}'+\zeta''\right) \nonumber\\
        && -\frac{G_{,X}}{a^{2}e^{2\alpha}}\phi'^{2}\left[\left(\mathcal{H}+\alpha'\right)\phi'-\phi''\right]+G_{,\phi}\phi'^{2} , {\label{zeta_eom}}\end{eqnarray}
respectively.
On the background level, by definition, $\alpha=\zeta=0$ is a set of solutions of equations (\ref{alpha_eom})-(\ref{zeta_eom}). Due to the absence of $\zeta$ term in the equations, one can explicitly verify that actually
    \begin{equation}
        \alpha=0,\qquad\qquad \zeta=c,
    \end{equation}
where $c$ is an arbitrary non-vanishing constant, is also a set of \emph{exact} solutions of the nonlinear equations (\ref{alpha_eom})-(\ref{zeta_eom}). This completes our proof that, on large scales, $\zeta$ is conserved non-perturbatively.
Note the equation for $\zeta$ (\ref{zeta_eom}) is \emph{linear} in $\zeta''$, which implies the system has a well-defined Cauchy initial-value problem.

\subsection{Generalized Galileon model}

The above analysis can be easily extended to the generalized Galileon model (\ref{action}). After some manipulation, the \emph{exact} large-scale Lagrangian takes the following form
    \begin{equation}{\label{L_ls_gg}}
        \mathcal{L} \approx e^{3\zeta}\left[ A(\zeta',\alpha) + \alpha' B(\zeta',\alpha) +\zeta'' C(\zeta',\alpha) \right]\,,
    \end{equation}
where $A,B,C$ are functions of $\zeta'$ and $\alpha$ as well as background quantities such as $a$ and $\mathcal{H}$, whose explicit expressions are given by
    \begin{eqnarray}
        A & = & \frac{1}{a^{2}}e^{-5\alpha}\Big[a^{6}e^{6\alpha}K^{ {(0)}}+6K^{ {(3)}}(\mathcal{H}+\zeta')^{2}\left(\phi'\right)^{2}\left((2\mathcal{H}-\zeta')\phi'-3\phi''\right)+\nonumber \\
         &  & a^{4}e^{4\alpha}\left(3(\mathcal{H}+\zeta')(\mathcal{H}+2\zeta')+3\mathcal{H}'+6Q^{ {(2)}}\left((\mathcal{H}+\zeta')(\mathcal{H}+2\zeta')+\mathcal{H}'\right)-K^{ {(1)}}\left((2\mathcal{H}+3\zeta')\phi'+\phi''\right)\right)\nonumber \\
         &  & -6a^{2}e^{2\alpha}(\mathcal{H}+\zeta')\left(-K^{ {(2)}}\phi'\left(\zeta'\phi'+\phi''\right)+3Q^{ {(3)}}\left(\left(3\zeta'(\mathcal{H}+\zeta')+2\mathcal{H}'\right)\phi'+(\mathcal{H}+\zeta')\phi''\right)\right)\Big],\label{A_ls}\\
        B & = & \frac{1}{a^{2}}e^{-5\alpha}\Big[18K^{ {(3)}}(\mathcal{H}+\zeta')^{2}\left(\phi'\right)^{3}-a^{4}e^{4\alpha}\left(3\left(1+2Q^{ {(2)}}\right)(\mathcal{H}+\zeta')-K^{ {(1)}}\phi'\right)\nonumber \\
         &  & +6a^{2}e^{2\alpha}(\mathcal{H}+\zeta')\phi'\left(9Q^{ {(3)}}(\mathcal{H}+\zeta')-K^{ {(2)}}\phi'\right)\Big],\label{B_ls}\\
        C & = & e^{-3\alpha}\left(3a^{2}e^{2\alpha}\left(1+2Q^{ {(2)}}\right)-36Q^{ {(3)}}(\mathcal{H}+\zeta')\phi'\right).\label{C_ls}
        \end{eqnarray}
We emphasize that in (\ref{A_ls})-(\ref{C_ls}), functions $K^{(0)}$ etc are their corresponding large-scale expressions by neglecting all spatial derivatives, e.g. $K^{(0)} = K^{(0)}(\frac{1}{2}\frac{e^{-2\alpha}}{a^{2}}\phi'^{2},\phi)$ etc.
Again, from (\ref{L_ls_gg}) it is explicit that $\zeta$ enters the Lagrangian (\ref{L_ls_gg}) only through an overall pre-factor $e^{3\zeta}$, under the constant shift $\zeta \rightarrow \zeta+c$, (\ref{L_ls_gg}) transform as
    \begin{equation}
        \mathcal{L} \rightarrow e^{3c}\mathcal{L}.
    \end{equation}
Based on the same analysis in the previous section, the system (\ref{L_ls_gg}) has an exact solution $\zeta=\text{const}$. This proves the conservation of $\zeta$ on large scales in the generalized Galileon model (\ref{action}).

\section{Conclusion}

The existence of a conserved curvature perturbation on super-Hubble scales is of particular importance in cosmology, which allows us relate the current observations with the theoretical predictions in the primordial era. The previous understanding of the conservation highly relies on the analysis of $k$-essence or perfect fluid models. Concerning the recently proposed ``generalized Galileon" models \cite{Deffayet:2011gz}, it is interesting to verify if this conservation law still holds.

In this work, we prove the existence of a fully nonlinear conserved curvature perturbation on large scales in Galileon scalar field models in two approaches. In the first approach, we show that the fluid corresponding to the Galileon field becomes perfect and barotropic on large scales, which are responsible to the conservation. The difference from $k$-essence model is that, besides the energy-momentum conservation, the Einstein equation must be employed in order to complete the proof of barotropy. As a more familiar and simpler formalism, we derive the fully non-perturbative action for the curvature perturbation $\zeta$ in Galileon model on large scales, and show that $\zeta=const$ is indeed an exact solution on large scales.

No need to say, both the two different approaches to prove the conservation are physically equivalent. The physical explanation of the conservation is  transparent: the Galileon models differ from $k$-essence model only with higher-order derivative terms. The effects of these higher-order derivative terms, however, vanish on large scales. This fact implies that, the Galileon field behaves like a $k$-essencial field on large scales. Thus, on large scales, the Galileon field must share the same properties with $k$-essencial field, including the conservation of curvature perturbation.
The aim of this work, is to show this conservation in a ``crystal clear" manner.

\acknowledgments
It is a pleasure to thank
David Langlois
and
Ignacy Sawicki
for intersting discussions. I am grateful to David Langlois for useful comments on a preliminary
version of this paper.
I am supported by ANR (Agence Nationale de la Recherche) grant ``STR-COSMO" ANR-09-BLAN-0157-01.

\appendix

\section{A general discussion on the entropy perturbation and barotropy}{\label{sec:baro_gen}}

In this appendix, we will make a slightly general discussion on the barotropy of the fluid corresponding to \emph{single} scalar field model (see \cite{Arroja:2010wy,Christopherson:2008ry} for related discussion in $k$-essence models). Especially, we will show that
if the energy density and the pressure can be written as a system of ``2-2" mapping:
    \begin{equation}{\label{rho_p_dphi_phi}}
        \rho=\rho(\dot{\phi},\phi,A_{(\rho)}), \qquad p=p(\dot{\phi},\phi,A_{(p)}),
    \end{equation}
where $A_{(\rho)}$ and $A_{(p)}$ formally represent terms which are irrelevant for our analysis in the specific problems, for our purpose there are higher order in spatial derivatives and can be neglected on large scales, the corresponding fluid always becomes barotrpic on large scales. Equivalently, the scalar perturbation becomes adiabatic  on large scales.

Without loss of generality, in the following we neglecte higher-order spatial derivative terms and consider $\rho=\rho(\dot{\phi},\phi)$ and $p=p(\dot{\phi},\phi)$.
Perturbing (\ref{rho_p_dphi_phi}) yields
    \begin{equation}
        D_a\rho=\rho_{\dot{\phi}}D_a\dot{\phi}+\rho_{\phi}D_a\phi,\qquad D_a p=p_{\dot{\phi}}D_a\dot{\phi}+p_{\phi}D_a\phi.
    \end{equation}
By solving $D_a\dot{\phi}$ in terms of $D_a\rho$ and $D_a\phi$, we get
    \begin{equation}{\label{p_rho_phi}}
        D_a p= c_{s}^{2}D_a\rho+\rho_{\phi}\left(\frac{p_{\phi}}{\rho_{\phi}}-\frac{p_{\dot{\phi}}}{\rho_{\dot{\phi}}}\right)D_a\phi,
    \end{equation}
where we define
    \begin{equation}
         c_{s}^{2}\equiv\frac{p_{\dot{\phi}}}{\rho_{\dot{\phi}}},
    \end{equation}
which has the meaning as ``propagating speed of scalar perturbation".

We can make a further variable transformation by introducing
    \begin{equation}{\label{cdp_phi_def}}
        \epsilon_a \equiv D_a\rho-\frac{\dot{\rho}}{\dot{\phi}}D_a\phi,
    \end{equation}
(\ref{cdp_phi_def}) can be explained as ``comoving density perturbation" in single scalar field model, since which becomes $D_a\rho$ in comoving frame of scalar field where $D_a\phi=0$. $\epsilon_a$ can also be explained as the ``relative entropy perturbation" between $\rho$ and $\phi$.
At this point, it is interesting to note if we alternatively introduce a ``relative entropy perturbation" between $\dot{\phi}$ and $\phi$ defined as
    \begin{equation}
        S_a(\dot{\phi},\phi) \equiv \frac{D_a\dot{\phi}}{\ddot{\phi}}-\frac{D_a\phi}{\dot{\phi}},
    \end{equation}
it is easy to show that
    \begin{equation}
        \epsilon_a = \rho_{\dot{\phi}}\ddot{\phi} S_a.
    \end{equation}
(\ref{cdp_phi_def}) can be used to replace $D_a\phi$ in (\ref{p_rho_phi}) and get
    \begin{equation}{\label{p_rho_cdp}}
        D_a p = c_{a}^{2}D_a\rho+\left(c_{s}^{2}-c_{a}^{2}\right) \epsilon_a,
    \end{equation}
where we define
    \begin{equation}
        c_a^2 \equiv \frac{\dot{p}}{\dot{\rho}}.
    \end{equation}
which has the meaning as ``adiabatic speed of sound". In terms of the non-adiabatic pressure, (\ref{p_rho_cdp}) can be written as an equivalent and more convenient form:
    \begin{equation}{\label{Gamma_epsilon}}
        \Gamma_a = \left(c_{s}^{2}-c_{a}^{2}\right) \epsilon_a.
    \end{equation}
Thus we get the very useful conclusion: if both the energy density and the pressure can be written as functions of $\dot{\phi}$ and $\phi$ as (\ref{rho_p_dphi_phi}), the following three quantities are proportional to each other:
    \begin{equation}
        \Gamma_a \propto \epsilon_a \propto S_a.
    \end{equation}

Now we are at the position to see under which condition the ``fluid" becomes barotropic, i.e. $D_a p \propto D_a \rho$ or equivalently $\Gamma_a = 0$. Typically, according to (\ref{Gamma_epsilon}), we are left with two possibilities in order to make $\Gamma_a$ vanish:
    \begin{itemize}
    \item One possibility is to make
    \begin{equation}{\label{baro_con1}}
        c_{s}^{2}=c_{a}^{2},
    \end{equation}
or equivalently
    \begin{equation}{\label{baro_con2}}
        \frac{\dot{p}}{\dot{\rho}}=\frac{p_{\phi}}{\rho_{\phi}}=\frac{p_{\dot{\phi}}}{\rho_{\dot{\phi}}}.
    \end{equation}
    If (\ref{baro_con1}) or (\ref{baro_con2}) is satisfied, the fluid becomes barotropic on \emph{all} scales.
    In general (\ref{baro_con1}) or (\ref{baro_con2}) put constraints on the functional forms of $\rho=\rho(\dot{\phi},\phi)$ and $\rho=\rho(\dot{\phi},\phi)$, and thus the structure of the scalar field theory. See \cite{Arroja:2010wy,Akhoury:2008nn,Christopherson:2008ry} for a recent discussion on the barotropy of the fluid corresponding to $k$-essence model and the constraint of exact barotropy on the functional form of $P(X,\phi)$.
    \item The other possibility  is to make
        \begin{equation}
            \epsilon_a=0.
        \end{equation}
        As is well-known in $k$-essence model, this indeed happens on \emph{large scales}, where the generalized Poisson equation yiedls $\epsilon_a \propto D_a(\nabla^2\Phi)\approx 0$, which yields the well-known conclusion that scalar perturbation of perfect fluid or $k$-essence model becomes adiabatic on large scales. As we will show in this work, this is still the case for Galileon model, although not as explicit as $k$-essence model \cite{Gordon:2000hv,Christopherson:2008ry,Langlois:2008mn}. See Appendix \ref{sec:supp_en} for a more general discussion on the suppression of entropy perturbation on large scales.
        Note on large scales $\epsilon_a=0$ implies $\dot{\phi} = \dot{\phi}(\phi)$, which can be viewed as the generalization of ``slow-roll" condition.
    \end{itemize}

In fact, we can consider the system which can be viewed as an ``$N$-$n$" mapping:
    \begin{equation}
        Q_I = Q_I (q_i), \qquad i=1,\cdots n,\quad I=1,\cdots N.
    \end{equation}
The relative entropy perturbation between any two $Q_I$'s is given by
    \begin{equation}
        S_{a}\left(Q_{I},Q_{J}\right) = \sum_{i,j} \frac{\dot{q}_{i}\dot{q}_{j}}{2\dot{Q}_{I}\dot{Q}_{J}}\left(\frac{\partial Q_{I}}{\partial q_{i}}\frac{\partial Q_{J}}{\partial q_{j}}-\frac{\partial Q_{J}}{\partial q_{i}}\frac{\partial Q_{I}}{\partial q_{j}}\right)S_{a}\left(q_{i},q_{j}\right),
    \end{equation}
which are linear combination of $n(n-1)/2$ relative entropy perturbation among $q_i$'s.
For our purpose, one may find the following ``$N$-2" mapping result be useful in practical calculations when $\{q_i\}= \{\dot{\phi},\phi\}$,
the
entropy perturbations between any of two $Q_I$'s are
    \begin{equation}
        S_{a}\left(Q_{I},Q_{J}\right) = \frac{\dot{\phi}\ddot{\phi}}{2\dot{Q}_{I}\dot{Q}_{J}}\left(\frac{\partial Q_{I}}{\partial\dot{\phi}}\frac{\partial Q_{J}}{\partial\phi}-\frac{\partial Q_{I}}{\partial\phi}\frac{\partial Q_{J}}{\partial\dot{\phi}}\right)S_{a}\left(\dot{\phi},\phi\right).
    \end{equation}
In this case, it is not surprise that all the relative entropy perturbations $S_a(Q_I,Q_J)$ are proportional to each other.
This is just what happens in our system. The key point is that, in single scalar field model, all other entropy modes are relative entropy modes among terms with higher-order spatial gradients and thus can be neglected on large scales by definition, except the relative entropy perturbation between $\dot{\phi}$ and $\phi$: $S_a(\dot{\phi},\phi)$ (although which itself is also suppressed by dynamical reason). Thus, $\Gamma_a$, $\epsilon_a$ and $\tilde{\epsilon}_a$ or any other quantity which has the meaning of relative entropy perturbation are just different but equivalent characterization of the same entropy perturbation of the system, and must be proportional to each other.

\section{Suppression of entropy perturbation on large scales}{\label{sec:supp_en}}

In this section, we will show that, as long as the energy density and the pressure of the scalar field model can be written in the form $\rho=\rho(\phi,\dot{\phi})$ and $P=P(\phi,\dot{\phi})$, the entropy/non-adiabatic perturbation in the corresponding system is suppressed on large scales.

To this end, we make use of the energy and momentum constraint:
    \begin{eqnarray}
        u^au^b G_{ab} \equiv \frac{1}{3}\Theta^2 + \frac{1}{2}\left( R^{(3)} - \sigma_{ab} \sigma^{ab}\right) & =&  \rho,\\
        u^b G_{bc} h^c_a \equiv D^b \sigma_{ba} - \frac{2}{3}D_a \Theta & = & -q_a,
    \end{eqnarray}
combination of which yields
    \begin{equation}{\label{cov_PE}}
        \frac{1}{2}D_{a}\left(R^{(3)}-\sigma_{ab}\sigma^{ab}\right)+\Theta D^{b}\sigma_{ba}=D_{a}\rho-\Theta q_{a}.
    \end{equation}
(\ref{cov_PE}) is the covariant generalization of Poission equation. The left-hand-side of (\ref{cov_PE}) is at least third-order in spatial derivatives and thus is suppressed on large scales in an almost FRW background, which implies the right-hand-side
    \begin{equation}
        \tilde{\epsilon}_a \equiv D_{a}\rho-\Theta q_{a},
    \end{equation}
is suppressed on large scales. The next task is to show $\tilde{\epsilon}_a$ is indeed an entropy perturbation, i.e. the characterization of the non-adiabatic perturbation in our system. To see this, we have to make use of the fact that, the energy-flow of the scalar field theory ($k$-essence, Galileons) arises only through spatial derivatives, i.e. its general structure is
    \begin{equation}
        q_a = \frac{q_{\phi}}{\dot{\phi}} D_a\phi + \frac{q_{\dot{\phi}}}{\ddot{\phi}} D_a\dot{\phi} +\mathcal{O}(D^2),
    \end{equation}
where $q_{\phi}$ and $q_{\dot{\phi}}$ are general coefficients depending on specific theories, whose explicit expressions are irrelevant to our following analysis.
Another important observation is, $q_a$ and $\rho$, $P$ must have the following constraint (neglecting higher order spatial derivative terms)
    \begin{equation}
         -q_{\phi}-q_{\dot{\phi}} \approx \rho+P,
    \end{equation}
which, although not obvious, can be verified explicitly through (\ref{EMT0}), (\ref{EMT1}), (\ref{EMT2}) and (\ref{EMT3}), or simply taking into account the functional form of the energy-momentum tensor{\footnote{Essentially, this is because the energy-momentum tensor of the scalar field is the variational derivative of a scalar Lagrangian with respect to the metric.}}.

Having the above relations in hand, we have
    \begin{eqnarray*}
        \tilde{\epsilon}_{a}-\epsilon_{a} & = & \frac{\dot{\rho}}{\dot{\phi}}D_{a}\phi-\Theta\left(\frac{q_{\phi}}{\dot{\phi}}D_{a}\phi+\frac{q_{\dot{\phi}}}{\ddot{\phi}}D_{a}\dot{\phi}+\mathcal{O}(D^{2})\right)\\
         & = & \frac{-\Theta\left(\rho+P\right)}{\dot{\phi}}D_{a}\phi-\Theta\frac{q_{\phi}}{\dot{\phi}}D_{a}\phi-\Theta\frac{q_{\dot{\phi}}}{\ddot{\phi}}D_{a}\dot{\phi}+\mathcal{O}(D^{2})\\
         & \equiv & \Theta q_{\dot{\phi}}S_{a}\left(\phi,\dot{\phi}\right)+\mathcal{O}(D^{2}),
        \end{eqnarray*}
where we used the energy-momentum constraint $\dot{\rho}+\Theta(\rho+P)= \mathcal{D}$ where $\mathcal{D}$ is defined in (\ref{curly_D_def}). The comoving density perturbation $\epsilon_a\equiv \dot{\rho}S_{a}\left(\rho,\phi\right)$ is just the entropy perturbation between $\rho$ and $\phi$, which is proportional to $S_a(\dot{\phi},\phi)$ on large scales since $\rho=\rho(\phi,\dot{\phi})$. Thus, we can conclude
    \begin{equation}
        \tilde{\epsilon}_a \propto S_a(\phi,\dot{\phi}),
    \end{equation}
i.e. is the characterization of the entropy perturbation in our system. As we emphasized before, the fact that our system is controlled (on large scales) by two degrees of freedom $\phi$ and $\dot{\phi}$ implies, all different entropy perturbations are proportional to each other, and are suppressed.

\section{Explicit expressions for $\mathcal{D}_{(\rho)}$, $\mathcal{D}_{(P)}$ and $\mathcal{D}_{(\ddot{\phi})}$}{\label{sec:explicit}}

The discussion of this work relies on the large-scale approximation. For completeness, here we collect various higher-order spatial derivative terms.
    \begin{eqnarray}
        \mathcal{D}_{\left(\rho\right)} & = & G_{,X}\Big[-\frac{1}{2}\dot{\phi}\left(D_{a}\phi D^{a}\phi\right)^{\cdot}+D_{a}XD^{a}\phi\nonumber \\
        &  & +\dot{\phi}^{2}\left(D^{a}D_{a}\phi+a^{a}D_{a}\phi\right)\Big]+G_{,\phi}D_{a}\phi D^{a}\phi,\label{D_rho}\end{eqnarray}
    \begin{eqnarray}
        \mathcal{D}_{\left(P\right)} & = & -G_{,X}\left[\frac{1}{2}\dot{\phi}\left(D_{a}\phi D^{a}\phi\right)^{\cdot}+\frac{1}{3}D_{a}XD^{a}\phi\right]\nonumber \\
        &  & +\left(\frac{1}{3}\tilde{B}-G_{,\phi}\right)D_{a}\phi D^{a}\phi,\label{D_P}\end{eqnarray}
and
    \begin{equation}{\label{D_ddotphi}}
        \mathcal{D}_{(\ddot{\phi})}=\frac{\rho_{,A_{\left(\rho\right)}}}{\rho_{,\dot{\phi}}}\dot{A}_{\left(\rho\right)}-\frac{3\rho_{,\Theta}}{2\Theta}\frac{\dot{\mathcal{R}}}{\rho_{,\dot{\phi}}}+\frac{1}{\rho_{,\dot{\phi}}}\left(1-\frac{3\rho_{,\Theta}}{2\Theta}\right)\left(\Theta\mathcal{D}_{\left(\rho\right)}+\Theta\mathcal{D}_{\left(P\right)}-\mathcal{D}\right).
    \end{equation}

\section{Relations among $\Gamma_a$, $\epsilon_a$, $\tilde{\epsilon}_a$}{\label{sec:rel_en}}

As have discussed in Appendix \ref{sec:baro_gen}, there are three quantities which are of particular importance in our analysis: the non-adiabatic pressure $\Gamma_a$, and two covariant generalizations of comoving density perturbation $\epsilon_a$ and $\tilde{\epsilon}_a$. In a perfect fluid or $k$-essence system, they are related by
    \begin{equation}
        \Gamma_a \approx (c_s^2 - c_a^2) \epsilon_a, \qquad \text{and}\qquad \epsilon_a \approx \tilde{\epsilon}_a.
    \end{equation}
In this appendix, we will derive the corresponding relations among the above three quantities in Galileon model (\ref{model}). We will frequently use the definition of relative entropy perturbation between any two quantities $S_a(X,Y)$ defined in (\ref{S_gen_def}).

To this end, it is useful to express them in terms of a basic quantity $S_a(\dot{\phi},\phi)$, which is the basic characterization of the relative entropy perturbation in our system.
We have derived the expression for $\rho$ and $P$ in terms of $\dot{\phi}$ and $\phi$ in (\ref{Da_rho})-(\ref{dot_P}), after some manipulations, we get:
        \begin{equation}
        \Gamma_a = \left(1-\frac{3\rho_{,\Theta}}{2\Theta}\right)^{-1}\frac{1}{\dot{\rho}}\left[C\dot{\phi}\ddot{\phi}S_{a}(\dot{\phi},\phi)+\mathcal{S}_{a}\right]
    \end{equation}
with
    \begin{equation}
        C= \rho_{,\phi}P_{,\dot{\phi}}-\rho_{,\dot{\phi}}P_{,\phi}+P_{,\ddot{\phi}}\left(\rho_{,\phi}\mathcal{A}_{,\dot{\phi}}-\rho_{,\dot{\phi}}\mathcal{A}_{,\phi}\right),
    \end{equation}
and
    \begin{eqnarray}
\mathcal{S}_{a} & = & P_{,\ddot{\phi}}\mathcal{A}_{,A_{(\ddot{\phi})}}\dot{A}_{(\ddot{\phi})}\left[\rho_{,\dot{\phi}}\ddot{\phi}S_{a}\left(A_{(\ddot{\phi})},\dot{\phi}\right)+\rho_{,\phi}\dot{\phi}S_{a}\left(A_{(\ddot{\phi})},\phi\right)\right]\nonumber \\
 &  & +P_{,A_{\left(P\right)}}\dot{A}_{\left(P\right)}\left[\rho_{,\dot{\phi}}\ddot{\phi}S_{a}\left(A_{\left(P\right)},\dot{\phi}\right)+\rho_{,\phi}\dot{\phi}S_{a}\left(A_{\left(P\right)},\phi\right)\right]\nonumber \\
 &  & +\left(P_{,\phi}+P_{,\ddot{\phi}}\mathcal{A}_{,\phi}\right)\dot{\phi}\left[\rho_{,A_{\left(\rho\right)}}\dot{A}_{\left(\rho\right)}S_{a}\left(\phi,A_{\left(\rho\right)}\right)+\frac{3\rho_{,\Theta}}{2\Theta}\dot{\mathcal{R}}S_{a}\left(\mathcal{R},\phi\right)\right]\nonumber \\
 &  & +\left(P_{,\dot{\phi}}+P_{,\ddot{\phi}}\mathcal{A}_{,\dot{\phi}}\right)\ddot{\phi}\left[\rho_{,A_{\left(\rho\right)}}\dot{A}_{\left(\rho\right)}S_{a}\left(\dot{\phi},A_{\left(\rho\right)}\right)+\frac{3\rho_{,\Theta}}{2\Theta}\dot{\mathcal{R}}S_{a}\left(\mathcal{R},\dot{\phi}\right)\right]\nonumber \\
 &  & +\dot{A}_{\left(\rho\right)}\rho_{,A_{\left(\rho\right)}}\left[P_{,\ddot{\phi}}\mathcal{A}_{,A_{(\ddot{\phi})}}\dot{A}_{(\ddot{\phi})}S_{a}\left(A_{(\ddot{\phi})},A_{\left(\rho\right)}\right)+P_{,A_{\left(P\right)}}\dot{A}_{\left(P\right)}S_{a}\left(A_{\left(P\right)},A_{\left(\rho\right)}\right)\right]\nonumber \\
 &  & +\frac{3\rho_{,\Theta}}{2\Theta}\dot{\mathcal{R}}\left[P_{,\ddot{\phi}}\mathcal{A}_{,A_{(\ddot{\phi})}}\dot{A}_{(\ddot{\phi})}S_{a}\left(\mathcal{R},A_{(\ddot{\phi})}\right)+P_{,A_{\left(P\right)}}\dot{A}_{\left(P\right)}S_{a}\left(\mathcal{R},A_{\left(P\right)}\right)\right].\label{mS_def}\end{eqnarray}
$\mathcal{S}_a$ is the summation of various relative entropy perturbations among terms which are higher-order in spatial derivatives, and can be completely neglected on large scales.
The comoving density perturbation $\epsilon_{a}$ defined in (\ref{cdp_def}) is given by
    \begin{equation}
        \epsilon_{a}=\left(1-\frac{3\rho_{,\Theta}}{2\Theta}\right)^{-1}\rho_{,\dot{\phi}}\ddot{\phi}S_{a}\left(\dot{\phi},\phi\right)+\mathcal{D}_{(\epsilon_{a})},
    \end{equation}
with
    \begin{equation}{\label{Dea}}
        \mathcal{D}_{(\epsilon_{a})}=\left(1-\frac{3\rho_{,\Theta}}{2\Theta}\right)^{-1}\left[\rho_{,A_{\left(\rho\right)}}\dot{A}_{\left(\rho\right)}S_{a}\left(A_{\left(\rho\right)},\phi\right)-\frac{3\rho_{,\Theta}}{2\Theta}\dot{\mathcal{R}}S_{a}\left(\mathcal{R},\phi\right)\right].
    \end{equation}
$\tilde{\epsilon}_a$ defined in (\ref{cdp_t_def}) is given by
    \begin{equation}
        \tilde{\epsilon}_{a}=\left[\left(1-\frac{3\rho_{,\Theta}}{2\Theta}\right)^{-1}\rho_{,\dot{\phi}}+\Theta G_{,X}\dot{\phi}^{2}\right]\ddot{\phi}S_{a}\left(\dot{\phi},\phi\right)+\mathcal{D}_{(\tilde{\epsilon}_{a})},
    \end{equation}
with
    \begin{eqnarray}
        \mathcal{D}_{(\tilde{\epsilon}_{a})} & = & \mathcal{D}_{(\epsilon_{a})}+\frac{1}{2}\Theta G_{,X}\dot{\phi}\left(D^{c}\phi D_{c}\phi\right)^{\cdot}S_{a}\left(\phi,\left(D^{b}\phi D_{b}\phi\right)\right) \nonumber\\
        &  & +\left[\mathcal{D}-\frac{\Theta}{3}\left(\tilde{B}D_{a}\phi D^{a}\phi+2G_{,X}D_{a}\phi D^{a}X\right)\right]\frac{D_{a}\phi}{\dot{\phi}}. \label{Dtea}\end{eqnarray}
Again, both $\mathcal{D}_{(\epsilon_a)}$ and $\mathcal{D}_{(\tilde{\epsilon}_a)}$ are higher order in spatial gradients, and should be neglected on large scales.

\section{Explicit expression for $T^{(3)}_{ab}$}{\label{sec:EMT3}}

The energy-momentum tensor corresponding to $\mathcal{L}^{(3)}$ takes the following form:
    \begin{eqnarray}
        T_{ab}^{(3)} & = & C_{1}g_{ab}+C_{2}\nabla_{a}\phi\nabla_{b}\phi+C_{3}\nabla_{a}X\nabla_{b}X+C_{4}\nabla_{(a}\phi\nabla_{b)}X+C_{5}\nabla_{a}\nabla_{b}\phi+C_{6}\nabla_{c}X\nabla^{c}\nabla_{(a}\phi\nabla_{b)}\phi\nonumber \\
         &  & +C_{7}\text{ }\nabla^{c}X\nabla_{c}\nabla_{d}\phi\nabla^{d}\nabla_{(a}\phi\nabla_{b)}\phi+C_{8}\nabla^{c}\nabla_{a}\phi\nabla_{b}\nabla_{c}\phi+C_{9}\nabla_{c}X\nabla_{(a}X\nabla^{c}\nabla_{b)}\phi\nonumber \\
         &  & +C_{10}\nabla_{c}\nabla_{d}\phi\nabla^{d}\nabla_{a}\phi\nabla^{c}\nabla_{b}\phi+\tau_{ab}^{(3)},\label{EMT3}
        \end{eqnarray}
where $C_1,\cdots,C_{10}$ are scalar coefficients given by:
    \begin{eqnarray}
        C_{1} & = & 6Q_{,\phi}^{(3)}\left[-\left(\square\phi\right)^{2}+\left(\nabla_{a}\nabla_{b}\phi\right)^{2}+2G_{ab}\nabla^{a}\phi\nabla^{b}\phi-RX\right]-6Q_{,\phi\phi}^{(3)}\left(\nabla_{a}X\nabla^{a}\phi-2X\square\phi\right)\nonumber \\
         &  & -6K_{,\phi}^{(3)}\left(X\left(\left(\nabla_{a}\nabla_{b}\phi\right){}^{2}-\left(\square\phi\right)^{2}\right)
         +2\nabla_{a}X\nabla^{a}X+2\square\phi\nabla_{a}X\nabla^{a}\phi\right)\nonumber \\
         &  & -3K^{(3)}\Big[\frac{2}{3}\left(\left(\Box\phi\right)^{3}-3\square\phi\left(\nabla_{a}\nabla_{b}\phi\right){}^{2}+2\left(\nabla_{a}\nabla_{b}\phi\right)^{3}\right)-2\square\phi R_{ab}\nabla^{a}\phi\nabla^{b}\phi\nonumber \\
         &  & -4R_{ab}\nabla^{a}X\nabla^{b}\phi+R\nabla_{a}X\nabla^{a}\phi+2R_{bdac}\nabla^{a}\phi\nabla^{b}\phi\nabla^{c}\nabla^{d}\phi\Big]\nonumber \\
         &  & -K_{,X}^{(3)}\left[3\left(\left(\square\phi\right)^{2}-\left(\nabla_{a}\nabla_{b}\phi\right){}^{2}\right)\nabla_{c}X\nabla^{c}\phi-6\nabla^{a}X\nabla^{b}X\nabla_{a}\nabla_{b}\phi+6\square\phi\nabla^{a}X\nabla_{a}X\right],\label{C1}
        \end{eqnarray}
        \begin{eqnarray}
        C_{2} & = & 6K_{,\phi}^{(3)}\left[\left(\square\phi\right)^{2}-\left(\nabla_{a}\nabla_{b}\phi\right){}^{2}\right]+6Q_{,\phi}^{(3)}R+6Q_{,\phi\phi}^{(3)}\square\phi-6K^{(3)}G_{ab}\nabla^{a}\nabla^{b}\phi\nonumber \\
         &  & +K_{,X}^{(3)}\left[\left(\square\phi\right)^{3}-3\left(\nabla_{a}\nabla_{b}\phi\right)^{2}\square\phi+2\left(\nabla_{a}\nabla_{b}\phi\right)^{3}\right],\label{C2}\\
        C_{3} & = & 12K_{,\phi}^{(3)}+6K_{,X}^{(3)}\square\phi,\label{C3}\\
        C_{4} & = & 6K^{(3)}R+6K_{,X}^{(3)}\left[\left(\square\phi\right)^{2}-\left(\nabla_{a}\nabla_{b}\phi\right)^{2}\right]+24K_{,\phi}^{(3)}\square\phi+12Q_{,\phi\phi}^{(3)},\label{C4}\\
        C_{5} & = & 12Q_{,\phi}^{(3)}\square\phi-12Q_{,\phi\phi}^{(3)}X-6K^{(3)}\left[\left(\nabla_{a}\nabla_{b}\phi\right)^{2}-\left(\square\phi\right)^{2}+R_{ab}\nabla^{a}\phi\nabla^{b}\phi\right]\nonumber \\
         &  & -2K_{,\phi}^{(3)}\left(6X\square\phi-6\nabla_{a}X\nabla^{a}\phi\right)+6K_{,X}^{(3)}\left(\nabla_{a}X\nabla^{a}X+\square\phi\nabla_{a}X\nabla^{a}\phi\right),\label{C5}\\
        C_{6} & = & -12K_{,X}^{(3)}\square\phi-24K_{,\phi}^{(3)},\label{C6}\\
        C_{7} & = & 12K_{,X}^{(3)},\label{C7}\\
        C_{8} & = & -12K^{(3)}\square\phi+12K_{,\phi}^{(3)}X-12Q_{,\phi}^{(3)}-6K_{,X}^{(3)}\nabla_{a}X\nabla^{a}\phi,\label{C8}\\
        C_{9} & = & -12K_{,X}^{(3)},\label{C9}\\
        C_{10} & = & 12K^{(3)},\label{C10}
        \end{eqnarray}
and $\tau_{ab}^{(3)}$ represents terms whose $a,b$-indices explicitly depend on or coupled to Riemannian tensors:
    \begin{eqnarray}
        \tau_{ab}^{(3)} & = & -12Q_{,\phi}^{(3)}\left(XR_{ab}+R_{acbd}\nabla^{c}\phi\nabla^{d}\phi+2\nabla_{(a}\phi R_{b)c}\nabla^{c}\phi\right)\nonumber \\
         &  & -6K^{(3)}\Big\{2\left[\left(\square\phi\nabla^{c}\phi+\nabla^{c}X\right)R_{c(a}+\nabla^{d}\nabla^{c}\phi\nabla^{e}\phi R_{ecd(a}-R_{cd}\nabla^{c}\phi\nabla^{d}\nabla_{(a}\phi\right]\nabla_{b)}\phi\nonumber \\
         &  & \qquad\qquad-\nabla_{c}X\nabla^{c}\phi R_{ab}+2\nabla^{c}\phi R_{c(a}\nabla_{b)}X\nonumber \\
         &  & \qquad\qquad-R_{c(ab)d}\nabla^{c}\phi\left(\square\phi\nabla^{d}\phi+2\nabla^{d}X\right)-2\nabla^{c}\phi\nabla^{e}\phi R_{edc(a}\nabla^{d}\nabla_{b)}\phi\Big\}.\label{tau3}
        \end{eqnarray}

In deriving the above expressions, we frequently used
    \[
        \left[\nabla_{b},\nabla_{c}\right]\nabla_{a}\phi=R_{\phantom{\rho}acb}^{d}\nabla_{d}\phi,\qquad \qquad \left[\nabla_{c},\nabla_{d}\right]\nabla_{a}\nabla_{b}\phi = R_{\phantom{\rho}bdc}^{e}\nabla_{a}\nabla_{e}\phi+R_{\phantom{\rho}adc}^{e}\nabla_{a}\nabla_{e}\phi,
    \]
as well as the Bianchi identity $\nabla_{a}R_{debc}+\nabla_{b}R_{deca}+\nabla_{c}R_{deab}=0$.

\section{3+1 decomposition of Riemannian quantities}{\label{sec:GC_rel}}

As we have seen in the bulk of this paper, besides the energy-momentum conservation equation, the Einstein equation must also be employed in order to complete the proof of conservation. Moreover, we frequently need to re-express the gravitational quantities in the equations in terms of fluid quantities. To this end, the ``3+1 decompositions" of the Einstein/Riemann/Ricci tensors, which are the counterparts of the corresponding decomposition of energy-momentum tensor $T_{ab}$ (\ref{EMT_dec}) and $\nabla_b u_a$ (\ref{Du_dec}), are needed.

These decompositions are the so-called Gauss/Codazzi/Ricci equations etc. One should keep in mind that they are purely kinetic equations, which are irrelevant to the underlying theory of gravitation. Some relevant equations which are used in this paper are summarized in the following{\footnote{Here we have expressed them in terms of $\Theta$, $\dot{\Theta}$, $a^a$, $h_{ab}$, $\sigma_{ab}$ etc for convenience. For their general and more compact expressions, see (e.g.) \cite{ADMref}.}:
    \begin{itemize}
        \item Contracted Gauss eqution
            \begin{equation}{\label{Gauss_con}}
                h_{a}^{c}h_{b}^{d}R_{cd}+h_{a}^{a'}h_{b}^{b'}u^{c}u^{d}R_{a'cb'd}={}^{3}R_{ab}+\frac{2}{9}\Theta^{2}h_{ab}+\frac{1}{3}\Theta\sigma_{ab}-\sigma_{ac}\sigma_{b}^{c},,
            \end{equation}
            where ${}^{3}R_{ab}$ is the spatial Ricci tensor constructed using $h_{ab}$.
        \item Contracted Codazzi equation
            \begin{equation}{\label{codazzi_con}}
                h^b_a u^c R_{bc} = D_{b}\sigma_{a}^{b} - \frac{2}{3}D_{a}\Theta.
            \end{equation}
        \item Ricci equation
            \begin{equation}{\label{Ricci_eq}}
                h_{a}^{a'}h_{b}^{b'}u^{c}u^{d}R_{a'cb'd}=-\frac{1}{9}\left(\Theta^{2}+3\dot{\Theta}\right)h_{ab}+a_{a}a_{b}+D_{a}a_{b}+\sigma_{ac}\sigma_{b}^{c}-\frac{1}{N}\pounds_{v}\sigma_{ab},
            \end{equation}
            where the last term is the Lie derivative with respect to $v^a\equiv N u^a$, with $N$ the lapse scalar. Under the large-scale approximation, only the first term on the right-hand-side in (\ref{Ricci_eq}) survives.
        \item Decomposition of Ricci scalar
            \begin{equation}{\label{Ricci_s_dec}}
                R = {}^{3}R+\frac{4}{3}\Theta^{2}+2\dot{\Theta}+\sigma_{ab}\sigma^{ab}-2a_{a}a^{a}-2D_{a}a^{a},
            \end{equation}
            where ${}^3R$ is the spatial Ricci scalar.
    \end{itemize}
Some other useful relations can be derived by combining (\ref{Gauss_con})-(\ref{Ricci_s_dec}), e.g.:
    \begin{eqnarray}
        u^{a}u^{b}R_{ab} & = & -\frac{1}{3}\Theta^{2}-\dot{\Theta}-\sigma_{ab}\sigma^{ab}+a_{a}a^{a}+D_{a}a^{a},\label{uuRicciT}\\
        h_{a}^{c}h_{b}^{d}R_{cd} & = & ^{3}R_{ab}+\frac{1}{3}\left(\Theta^{2}+\dot{\Theta}\right)h_{ab}-a_{a}a_{b}-D_{a}a_{b}+\frac{1}{3}\Theta\sigma_{ab}-2\sigma_{ac}\sigma_{b}^{c}+\frac{1}{N}\pounds_{v}\sigma_{ab}.\label{hhRicciT}
        \end{eqnarray}
Note in the above expressions $\omega_{ab}=0$ is assumed, which is the case when $u^a$ is hypersurface orthogonal.



\end{document}